\documentclass[aps,prd,twocolumn,superscriptaddress,groupedaddress, nofootinbib]{revtex4}  % for review and submission
\usepackage{graphicx}% Include figure files
\usepackage{dcolumn}% Align table columns on decimal point
\usepackage{bm}% bold math
\usepackage{hyperref}% add hypertext capabilities
\usepackage{float}
\usepackage{epsfig,graphics,colordvi}
\usepackage{amsmath}
\usepackage{amssymb}
\usepackage{mciteplus}
\usepackage{dsfont}

\usepackage{graphicx}
\usepackage[english]{babel}
\usepackage{sidecap}
\usepackage{multirow}
\usepackage{afterpage}

\usepackage{mathrsfs}

\usepackage{color}
\usepackage[usenames,dvipsnames]{xcolor}
\usepackage{bbold}

\usepackage{hyperref}

\usepackage{hyphenat}

\newcommand{\tr}{\operatorname{tr}}
\newcommand{\te}{\text}
\newcommand{\nn}{\nonumber}

\newcommand{\Lag}{\mathscr{L}}
\newcommand{\order}{\mathcal{O}}
\newcommand{\chpt}{{\chi\te{PT}}}
\newcommand{\bra}{\left\langle}
\newcommand{\ket}{\right\rangle}

\newcommand{\Lvec}{\Lag_{\te{vec}}}

\newcommand{\Llin}{\Lag_{\te{lin}}}
\newcommand{\Lch}{\Lag_\chpt}

\newcommand{\Um}{\mathcal{U}}

\newcommand{\mv}{m_V}

\newcommand{\lM}{\mathring M}
\newcommand{\mR}{\lM_{\! R}}
\newcommand{\mP}{\lM_{\! P}}
\newcommand{\mpi}{\lM_{\pi}}
\newcommand{\mk}{\lM_{\! K}}
\newcommand{\meta}{\lM_{\eta}}

\newcommand{\adb}{\allowdisplaybreaks}

\DeclareMathOperator{\SelfEn}{\operatorname{\Sigma}}

\newcommand{\ie}{\textit{i.e.}, }
\newcommand{\eg}{\textit{e.g.}, }

\begin{document}

\title{Contributions of loops with dynamical vector mesons to masses and decay constants of pseudoscalar mesons and their quark mass dependence}

%\titlerunning{Renormalisation of the $\chpt$-low-energy coupling constants with vector-meson loops}        % if too long for running head

\author{Carla Terschl\"usen and Stefan Leupold}  

\affiliation{Institutionen f\"or fysik och astronomi, Uppsala Universitet, Box 516, 75120 Uppsala, Sweden}
%
%\authorrunning{C.\ Terschl\"usen, S.\ Leupold}
\begin{abstract}
	The contributions of one-loop diagrams with dynamical vector mesons to masses and decay constants of pseudoscalar mesons are determined. Hereby, a relativistic Lagrangian for both the pseudoscalar-meson octet and the vector-meson nonet is used. The vector mesons are given in the antisymmetric tensor representation. Both the differences between static and dynamical vector mesons and the differences between calculations with and without vector mesons are studied as functions of the light quark mass.
\end{abstract}
%
%\pacs{}
%
\maketitle

%\tableofcontents

\section{Introduction}
\label{sec:Intro}

How important are vector mesons for low-energy QCD? Conceptually there is a clear answer to this question: For small enough 
momenta {\em and small enough quark masses} chiral perturbation theory ($\chi$PT) \cite{Weinberg:1978kz, Gasser:1983yg, Gasser:1984gg, Scherer:2002tk,Scherer-Schindler} constitutes the low-energy 
effective field theory of QCD. The degrees of freedom (DOF) of $\chi$PT are the pseudoscalar quasi-Goldstone bosons that 
emerge from the 
spontaneous breaking of the approximate chiral symmetry. For the two lightest quark flavours these bosons are the pions, 
for three flavours the pions, kaons and the $\eta$-meson. The masses of the quasi-Goldstone bosons are related to the 
non-vanishing quark masses. The influence of vector mesons, like of all other non-Goldstone-boson states, is encoded in the 
low-energy constants of $\chi$PT \cite{Gasser:1983yg, Ecker:1988te, Donoghue:1988ed}. 

Conceptually this is a clear-cut scheme, but quantitatively one wants to know how fast or slow the series expansion defined by the power counting of $\chi$PT converges. It is clear that the expansion scheme becomes the worse, the smaller the gap is between the masses of the active DOF and the not explicitly included states. In reality there is a comfortably large mass gap between the pions and all other hadronic states \cite{PDG2015}. However, already when one considers the three lightest quark flavours, then the mass gap is not tremendously large between the kaons and the $\eta$-meson as active DOF on the one hand and the vector mesons on the other\footnote{In this line of reasoning one might also include the sigma meson. But in the present exploratory work we restrict ourselves to the vector mesons.}.
In addition, a significant number of lattice-QCD calculations uses quark masses that are larger than in reality \cite{Gattringer:2010zz}. 
To relate these lattice-QCD results to the real world requires that the extrapolation to physical quark masses is theoretically
well understood. Also here $\chi$PT becomes instrumental with its systematic expansion in powers and logarithms of the 
masses of the quasi-Goldstone bosons \cite{Sharpe:2006pu}. But also this line of reasoning brings us back to the 
question about the quality of the $\chi$PT expansion in a real or lattice ``world'' where the quark masses are small, 
but not very small, in other words for the case where the quasi-Goldstone bosons are lighter, but not very much lighter 
than other DOF. Vector mesons constitute an important example for such not very heavy other DOF.

As already discussed, vector mesons appear in the low-energy constants of $\chi$PT. Essentially vector-meson propagators 
are expanded in powers of momenta over vector-meson masses \cite{Gasser:1983yg, Ecker:1988te,Donoghue:1988ed}. In this way
vector mesons become ``static''. In the present work we address the question how quantitatively different the effects 
from dynamical versus static vector mesons might be. To be specific we will calculate the masses and decay constants of the 
quasi-Goldstone bosons and study their dependence on the quark masses for the two cases where vector mesons are included 
as static or dynamical DOF in one-loop contributions. 

To answer such a quantitative question requires a quantitatively reasonable input. Fortunately vector mesons are 
phenomenologically rather well explored based on their prominent appearance in interactions between hadrons and 
electromagnetism \cite{Sakurai:1969}. In particular the coupling of the $\rho$-meson to photons (to external vector sources) 
and to a pair of pions has been studied in great detail. The three-flavour version of the latter interaction will provide our 
input for the one-loop calculations. We would like to admit right away that the present work is a first exploratory study. 
Therefore we do not aim at a systematic inclusion of vector mesons. Instead we use a particular, phenomenologically well motivated
model Lagrangian for the vector mesons with the interaction that we regard as most important: the $V$-$P$-$P$ interaction where
$V$/$P$ denotes a vector/pseudoscalar meson. Note that there are other effects and interaction types connected to 
vector mesons. In particular, in the present work we neither include the mass splitting within the vector-meson multiplet nor 
the $V$-$V$-$P$ interaction, see, e.g., \cite{diplCarla, Terschlusen:2012xw, Terschlusen:2013iqa} and references therein.  

Dealing with loops requires renormalisation. This issue, the divergence structure of loops including dynamical vector mesons, 
has been addressed in our previous work \cite{inf}. We also refer to this paper for a much more detailed discussion of scale 
separation and the importance of vector mesons. Here we can utilise the results from \cite{inf} and study in the 
present work the finite parts of one-loop diagrams including vector mesons as they contribute to the two-point functions 
of the pseudoscalar mesons. The present work focuses on two comparisons, for both cases as a function of the quark masses: 
1.\ We compare loops with static vector mesons and loops with dynamical vector mesons. 2.\ We compare loops with (dynamical) 
vector mesons and pure $\chi$PT loops. 

For the first comparison we start with the observation that loops with (static) vector mesons make their (indirect) appearance 
in $\chi$PT at next-to-next-to-leading order (N$^2$LO) \cite{Amoros:1999dp}. At this order one-loop diagrams with vertices from the 
next-to-leading order (NLO) Lagrangian contribute. In turn the corresponding NLO low-energy constants are influenced by 
vector mesons \cite{Gasser:1983yg, Ecker:1988te,Donoghue:1988ed} (and other mesons, but here we focus on vector mesons). With our comparison we study how important the difference between static and dynamical vector mesons actually is. We recall that we study this issue as a function of the quark masses.

For the second comparison we imagine the following two ``microscopic'' Lagrangians: one with pseudoscalar mesons only, one 
containing vector mesons in addition. At one-loop accuracy we fully integrate out the vector mesons 
for the second Lagrangian \cite{inf}
and in any case the pseudoscalar fluctuations \cite{Gasser:1984gg}. Concerning the divergence structure we restrict our 
attention to the chiral orders $Q^2$ and $Q^4$. We will see below that this is sufficient to address the two-point functions 
of pseudoscalar mesons, the topic of the present work. Starting with or without vector mesons we obtain a low-energy effective 
action. In the spirit of effective field theories, i.e.\ assuming that there is {\em an} effective field theory, not several
ones, it should be clear that at any chiral order the difference between the two scenarios 
(starting with or without vector mesons) can only reside in different values of the low-energy constants. Now suppose that we 
adjust the low-energy constants of the chiral orders $Q^2$ and $Q^4$ such that there is no difference between the two scenarios 
for observables up to (including) order $Q^4$. There is still a difference left between the two scenarios: The effective action 
obtained from starting with vector mesons contains finite non-local terms (logarithms) that depend on the vector-meson mass 
and the masses of the quasi-Goldstone bosons. If one expands these terms in powers of $Q$ over the vector-meson mass they 
start to contribute at chiral order $Q^6$. By not expanding these terms but keeping the full analytic structure we can explore 
how quantitatively important such terms are for a formal $Q^4$ calculation of the masses and decay constants of the 
quasi-Goldstone bosons. Again we address this question as a function of the quark masses.

Since all our calculations are performed as a function of the quark masses, a natural application of our work is a comparison 
to lattice-QCD results \cite{Aoki:2013ldr, Bazavov:2011fh, Carrasco:2014cwa}. In this present exploratory study we refrain from a direct comparison. The reason is that there are additional effects that one might want to consider when comparing to lattice results, in particular finite-volume
effects \cite{Bijnens:2014dea, Bijnens:2015dra}. This is beyond the scope of the present work, but we regard 
our results for the quark-mass dependence of the influence of vector-meson loops interesting enough to present them here. 

This article is structured in the following way: First, the necessary definitions are introduced and the results from \cite{inf} which are needed for this article are summarised (section \ref{sec:gen}). Then, one-loop contributions to pseudoscalar masses (section \ref{sec:masses}) and decay constants are determined (section \ref{sec:decay-const}). In section \ref{sec:Numerics}, these contributions are evaluated numerically. A summary is given in the last section.

\section{General considerations and renormalisation of low-energy constants of $\chpt$} \label{sec:gen}

We start with the Lagrangian used in \cite{inf} where the renormalisation aspects of one-loop calculation with dynamical vector mesons have been addressed. While \cite{inf} provides a feasibility test of the beyond-tree-level calculations with vector mesons, the present work studies the influence of loops with vector mesons on physical observables.

Within this article, calculations are done for the pseudoscalar octet only, the singlet is not taken into account (see also \cite{Terschlusen:2012xw, inf} for possible extensions). The pseudoscalar octet is included in the matrix field $U {:=} \exp(i\Phi/F)$ and given by
\begin{align}
	\Phi = \begin{pmatrix} \pi^0 + \frac{1}{\sqrt{3}}\eta_8 & \sqrt{2} \pi^+ & \sqrt{2} K^+ \\
							\sqrt{2} \pi^- & -\pi^0 + \frac{1}{\sqrt{3}} \eta_8 & \sqrt{2} K^0 \\
							\sqrt{2} K^- & \sqrt{2} \bar{K}^0 & -\frac{2}{\sqrt{3}} \eta_8
			\end{pmatrix}.
\end{align}
The leading-order-(LO) and NLO-$\chpt$ Lagrangians are denoted as in \cite{Gasser:1984gg},
\begin{align}
	& \Lch^{\te{LO}} = \frac{1}{4} F^2 \left\{ \bra D_\mu U^\dagger \, D^\mu U \ket + \bra \chi U^\dagger + \chi^\dagger U \ket \right\}  , \nn \adb \\
	& \Lch^{\te{NLO}} = L_1 \bra D_\mu U^\dagger \, D^\mu U \ket^2 + L_2 \bra D_\mu U^\dagger \, D_\nu U \ket^2 \nn \adb \\
	& \phantom{\Lch^{\te{NLO}} =} + L_3 \bra (D_\mu U^\dagger \, D^\mu U)^2 \ket \nn \adb \\
	& \phantom{\Lch^{\te{NLO}} =} + L_4 \bra D_\mu U^\dagger \, D^\mu U \ket \bra \chi^\dagger U + \chi U^\dagger \ket \nn \adb \\
	&  \phantom{\Lch^{\te{NLO}} =} + L_5 \bra (D_\mu U^\dagger \, D^\mu U) (\chi^\dagger U + U^\dagger \chi) \ket \nn \adb \\
	& \phantom{\Lch^{\te{NLO}} =} + L_6 \bra \chi^\dagger U + \chi U^\dagger \ket^2 \nn \adb \\
	&  \phantom{\Lch^{\te{NLO}} =} + L_7 \bra \chi^\dagger U - \chi U^\dagger \ket^2 + L_8 \bra \chi^\dagger U \chi^\dagger U + \chi U^\dagger \chi U^\dagger \ket \nn \adb \\
	&  \phantom{\Lch^{\te{NLO}} =} - iL_9 \bra F_R^{\mu\nu} D_\mu U \,  D_\nu U^\dagger + F_L^{\mu\nu} D_\mu U^\dagger \, D_\nu U \ket \nn \adb \\
	&  \phantom{\Lch^{\te{NLO}} =} + L_{10} \bra U^\dagger F_R^{\mu\nu} U F^L_{\mu\nu} \ket \nn \adb \\
	&  \phantom{\Lch^{\te{NLO}} =} + H_1 \bra F^R_{\mu\nu} F_R^{\mu\nu} + F^L_{\mu\nu} F_L^{\mu\nu} \ket + H_2 \bra \chi^\dagger \chi \ket. \label{eq:ChPTLagr}
\end{align}
Thereby, $\chi {:=} 2 B_0 (s + ip)$ including the external scalar and pseudoscalar sources $s$ and $p$, respectively. If the external fields are switched off, $\chi {=} 2 B_0 \mathcal{M} {:=} 2 B_0 \text{diag} (m,m,m_s)$ with an averaged up- and down-quark mass $2m {=} m_u + m_d$ and the mass $m_s$ of the strange quark. Furthermore,\footnote{Note that the covariant derivative $D_\mu$ is defined depending on the field it is acting on and acts differently on $U$, $U^\dagger$ and the vector field $V$.}
\begin{align}
	& D_\mu U := \partial_\mu U - i F^R_\mu U + i U F^L_\mu\,, \nn \adb \\
	& D_\mu U^\dagger := \partial_\mu U^\dagger + i U^\dagger F_R^\mu - i F_L^\mu U^\dagger \,, \nn \adb \\
	& F_{\mu\nu}^{R/L} := \partial_\mu F_\nu^{R/L} - \partial_\nu F_\mu^{R/L} - i \left[ F_\mu^{R/L}, F_\nu^{R/L} \right].
\end{align} 
As in \cite{inf}, the Lagrangian for vector mesons is restricted to two interaction terms for vector mesons $V$, pseudoscalar mesons $P$ and an external vector source $v$, a $V$-$2P$ and a $V$-$v$ interaction term. These interaction terms are the most important ones since they describe among other a $\rho$-$2\pi$ interaction and the transition of a vector meson into a photon \cite{Sakurai:1969}. The Lagrangian for vector mesons used in this article is given as \cite{Lutz:2008km, Terschlusen:2012xw, inf}
\begin{align}
	&\Lvec = \Lag_{\te{free}} + \Lag_{\te{lin}}, \nn \adb \\
	&\Lag_{\te{free}} = -\frac{1}{4} \bra D^\mu V_{\mu\nu} \, D_{\rho}V^{\rho\nu} \ket + \frac{1}{8} \, \mv^2 \bra V_{\mu\nu} V^{\mu\nu} \ket , \nn \adb \\
	&\Lag_{\te{lin}} = \frac{1}{2} i f_V h_P \bra \Um_\mu V^{\mu\nu} \Um_\nu  \ket + \frac{1}{2} f_V \bra V^{\mu\nu} f^+_{\mu\nu} \ket 	\label{eq:Lvec}
\end{align}
with $\bra A \ket := \tr(A)$, the parameters $f_V$ and $h_P$ and an approximated common vector-meson mass $\mv {=} 776 \, \te{MeV}$. Furthermore,  
\begin{align}
	& D_\mu V_{\alpha\beta} := \partial_\mu V_{\alpha\beta} + \left[\Gamma_\mu, V_{\alpha\beta} \right], \nn \adb\\
	& \Gamma_\mu := \frac{1}{2} \left( \left[u^\dagger, \partial_\mu u \right] - i u^\dagger F_\mu^R u + i u F^L_\mu u^\dagger \right), \nn \adb \\
	& \Um_\mu := \frac{1}{2} u^\dagger D_\mu U u^\dagger = -\frac{1}{2} u (D_\mu U)^\dagger u, \nn \adb \\
	& f_{\mu\nu}^\pm := \frac{1}{2} \left(u F_{\mu\nu}^L u^\dagger \pm u^\dagger F_{\mu\nu}^R u \right), \nn \adb \\	
	& U = u^2\,.
\end{align}
Hereby, the external vector and axialvector sources $v_\mu$ and $a_\mu$, respectively, are included in $F_\mu^{R/L} := v_\mu \pm a_\mu$. The vector mesons are given in antisymmetric tensor representation \cite{Gasser:1983yg, Ecker:1988te, diplCarla, Terschlusen:2012xw, Terschlusen:2013iqa, inf, Lutz:2008km, Ecker:1989yg} and collected in the nonet matrix
\begin{align}
	V_{\mu\nu} = \begin{pmatrix} \rho^0_{\mu\nu} + \omega_{\mu\nu} & \sqrt{2} \rho^+_{\mu\nu} & \sqrt{2} K^+_{\mu\nu} \\
						 \sqrt{2} \rho^-_{\mu\nu} & -\rho^0_{\mu\nu} + \omega_{\mu\nu} & \sqrt{2}K^0_{\mu\nu} \\
						 \sqrt{2} K^-_{\mu\nu} & \sqrt{2} \bar{K}^0_{\mu\nu} & \sqrt{2} \phi_{\mu\nu} \end{pmatrix} \! .
\end{align}

In $\chpt$ up to order $Q^4$, only the low-energy constants of the NLO-$\chpt$ Lagrangian are renormalised by pseudoscalar loops \cite{Gasser:1983yg, Gasser:1984gg}. If, however, loops with vector mesons are taken into account, an additional infinite one-loop contribution proportional to the kinetic term $\bra D_\mu U^\dagger \, D^\mu U \ket$ in the LO Lagrangian $\Lch^{\te{LO}}$ will be produced \cite{inf}. Therewith, the wave-function normalisation (wfn) constant $F$ as a coefficient of the kinetic term has to be renormalised as well. The renormalised wfn constant $F_r$ is given by \cite{inf}
\begin{align}
	&F_r^2 = F^2 + \frac{\varphi \, \mv^2}{F_r^2} \bar{\lambda}\, , \ \ \varphi := -\, \frac{9 \, f_V^2 h_P^2}{16}\, , \adb \nn \\
	&\bar{\lambda} := \frac{1}{16\pi^2} \left( \frac{1}{\varepsilon} + \operatorname{\Gamma}'\!(1) - 1 - \log (4\pi)  \right). \label{eq:Ren-F}
\end{align}
Thereby, the loop calculations are carried out with dimensional regularisation in $(4+2\varepsilon)$ dimensions. The infinities are identified via a modified MS-bar scheme according to \cite{Gasser:1983yg, Gasser:1984gg}. Since only the kinetic term but not the mass term in $\Lch^{\te{LO}}$ is renormalised directly, the renormalised LO-$\chpt$ Lagrangian is given by
\begin{align*}
	\Lch^{\te{LO}} = \frac{1}{4} F_r^2 \bra D_\mu U^\dagger D^\mu U \ket + \frac{1}{4} F^2 \bra \chi^\dagger U + \chi U^\dagger \ket. 
\end{align*}
Note that the combination $F^2 B_0 m_q$ remains finite where $m_q$ denotes a quark mass. Based on the renormalised Lagrangian, the field $U$ has to be redefined as
\begin{align}
	U = \exp(i\Phi/F) \mapsto U = \exp(i\Phi/F_r).
\end{align}
Therewith, the bare masses of the pseudoscalar mesons are equal to
\begin{align}
	&\mpi^2 = \frac{F^2}{F_r^2} \, 2 B_0 m \, , \  \mk^2 = \frac{F^2}{F_r^2} \, B_0 (m+m_s) \, , \nn \\
	&\meta^2 = \frac{F^2}{F_r^2} \, \frac{2}{3} B_0 (m+2m_s) = \frac{1}{3} \left( 4 \mk^2 - \mpi^2 \right) \label{eq:Def-bare-mass}
\end{align}
differing from the bare masses in pure $\chpt$ by a factor of $F^2/F_r^2$. When we explore the quark-mass dependence of our results, we study in practice the variations as a function of the bare pion mass $\mpi$. 

Furthermore, the contributions from loops with vector mesons to the low-energy constants $L_i$ of the NLO-$\chpt$ Lagrangian depend on the renormalised wfn constant $F_r^2$ via \cite{inf}
\begin{align}
	&L_i^r = L_i + \left( \frac{1}{2} \Gamma_i + \frac{\Lambda_i}{F_r^2} \right) \bar{\lambda}. \label{eq:Ren-const}
\end{align}
Hereby, $\Gamma_i$ denote the renormalisation constants from pure $\chpt$  \cite{Gasser:1984gg}\footnote{Note that the parameter $\bar{\lambda}$ as defined in Eq.\ \eqref{eq:Ren-F} is twice the corresponding parameter in \cite{Gasser:1984gg} yielding coefficients $\Gamma_i/2$.} and $\Lambda_i$ the renormalisation constants from loops with vector mesons. The values for the low-energy constants and the corresponding renormalisation constants relevant for the calculations within this article are listed in Tab. \ref{tab:chpt-le-constants}.
\begin{table}[h]
\caption{Phenomenologically determined values at $\mu {=} \mv$ for low-energy constants \cite{Bijnens:2014lea} needed for the calculations within this article and their respective renormalisation constants. $\Gamma_i$ denote the renormalisation constants from pure $\chpt$ \cite{Scherer:2002tk}, $\Lambda_i$ those from loops with vector mesons. $f_V$ and $h_P$ are parameters of the vector Lagrangian $\Llin$ (cf.\ Eq.\ \eqref{eq:Lvec}).}
\label{tab:chpt-le-constants}
\begin{tabular}{c| c |c |c}
 low-energy & phenom. value  & \multirow{2}{*}{$\Gamma_i$} & \multirow{2}{*}{$\Lambda_i /(f_V^2 h_P^2)$} \\
 constant & $[10^{-3}]$ & \\ \hline 
 & & \\[-0.7em]
 $L_4$ & $\phantom{-}0.0 \pm 0.3$ & $-\frac{1}{8}$ & $-\frac{3}{256}$ \\[0.3em]
 $L_5$ & $\phantom{-} 1.2 \pm 0.1$ &  $-\frac{3}{8}$ & $-\frac{9}{256}$ \\[0.3em]
 $L_6$ & $\phantom{-} 0.0 \pm 0.4$ & $\frac{11}{144}$ & $\phantom{-}0$ \\[0.3em]
 $L_7$ & $-0.3 \pm 0.2$ & $0$ & $\phantom{-}\frac{1}{256}$ \\[0.3em]
 $L_8$ & $\phantom{-} 0.5 \pm 0.2$ & $\frac{5}{48}$ & $-\frac{3}{256}$
\end{tabular}
\end{table} 
\section{One-loop contributions to masses of pseudoscalar mesons} \label{sec:masses}

The mass of a pseudoscalar meson is one physical observable used to study the influence of loops with vector mesons within this article. In this section, masses of pseudoscalar particles are determined generally, the numerical results are discussed in section \ref{sec:Numerics}.

\begin{figure}[b]
 	\begin{center} $
	 \begin{array}{ccc}
	 \includegraphics[trim = 0 22.5 0 -22.5, width=0.12\textwidth]{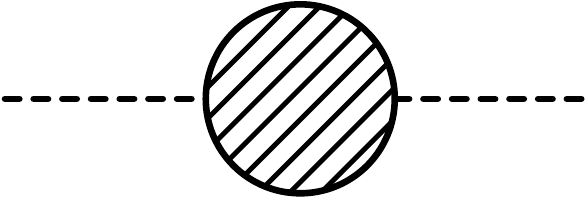} \vspace{1em} &
	\large{\ = \ } &  \hspace{-2em}  
	 \includegraphics[trim = -0.5em -5 0.5em 5 , width=0.08\textwidth]{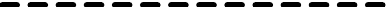}  \hspace{0.5em} \large{\ + \ } \hspace{0.5em}  \includegraphics[trim = 0 22.5 0 	-22.5,  width=0.12\textwidth]{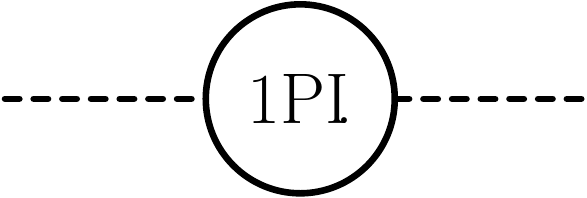} \\ 
	 & & \hspace{-.5em} \large{\ + \ } \hspace{0.5em} 
	 \includegraphics[trim = 0 22.5 0 -22.5, width=0.18\textwidth]{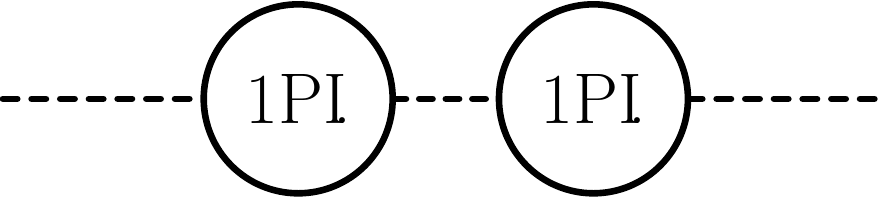} \hspace{0.5em}  \large{\ + \ } $\ldots$   
	 \end{array} $
	% width = x, 2/3*x, x, 1.5*x 
	%[trim = l b r t]
	 \caption{Propagator given as the sum of diagrams including the (irreducible) self energy. The dashed circle denotes the full contribution, ``1PI" denotes one-particle irreducible contributions.}
 	\label{fig:PropDiag}
 	\end{center}
	\end{figure}

In general, the mass $M$ of a particle is defined as the position of the pole of its propagator $\Delta$ as a function of the squared momentum $p^2$ of the incoming particle, \ie
\begin{align*}
	\Delta (p^2 = M^2)^{-1} \equiv 0.
\end{align*}
In LO $\chpt$, the propagator for a given pseudoscalar meson reads as
\begin{align}
	\Delta(p^2) = \frac{1}{p^2 - \lM^2+i0^+} \label{eq:prop-allg}
\end{align}
with the bare mass $\lM$ of the pseudoscalar meson as defined via the mass term in the LO-$\chpt$ Lagrangian $\Lch^{\te{LO}}$. Hence, the LO mass of a pseudoscalar meson in pure $\chpt$ is equal to its bare mass. If higher-order contributions and/or non-trivial LO contributions are included, the propagator can be expressed as an infinite sum of diagrams (see Fig.\ \ref{fig:PropDiag}).
%
%\afterpage{
%}
%
Defining the self energy $-i \SelfEn(p^2)$ as the sum of all one-particle-irreducible diagrams at a given chiral order, the full propagator at this order can be expressed as a geometric series \cite{Peskin:1995ev},
\begin{align}
	i \Delta(p^2) &= \frac{i}{p^2 - \lM^2+i0^+ } + \frac{ i \left[-i \SelfEn(p^2)\right]i }{( p^2 - \lM^2+i0^+)^2} + \ldots \nn \\
	&= \frac{i}{p^2 - \lM^2 - \SelfEn(p^2) +i0^+}\,  .
\end{align}
The (full) mass $M$ of a particle is the pole of the (full) propagator and, thus, defined via the mass equation
\begin{align}
	M^2 - \lM^2 - \SelfEn(M^2) = 0\,. \label{eq:mass-eq}
\end{align}

\begin{figure}[b]
	\begin{center} $
	\begin{array}{llcc}
		\text{(I)} & \text{pure } \chpt \text{:} \vspace{1em} &
		\hspace{0.5em} \includegraphics[trim = 0 3 0 -3, width=0.08\textwidth]{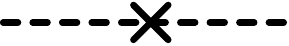} &
		\hspace{-1.5em} \includegraphics[trim = 0 -1 0 1, width=0.08\textwidth]{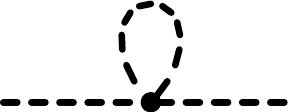}	\\
		\text{(II)} & \multicolumn{3}{l}{\text{loops with vector mesons:}} \\[0.5em] 
		& \hspace{0.5em} \multirow{2}{*}{\includegraphics[trim = 0 -1 0 1, width=0.08\textwidth]{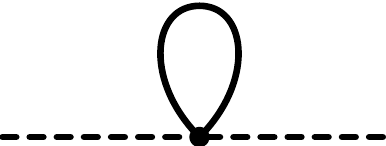}} &
		\hspace{0.5em} \multirow{2}{*}{\includegraphics[trim = 0 11 0 -11, width=0.12\textwidth]{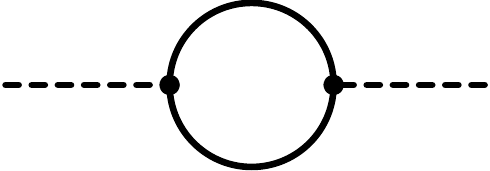}} &
		\hspace{0.5em} \multirow{2}{*}{\includegraphics[trim = 0 11 0 -11, width=0.12\textwidth]{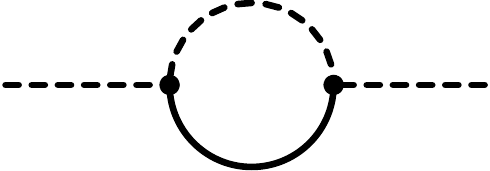}} \\[2em]		
	\end{array} $
% width = x, x, 3/2x, 3/2x	
	\caption{Contributing one-loop diagrams to the self energy $-i\SelfEn$ from pure $\chpt$ at $\order(Q^4)$ (I) and from loops including vector mesons (II). A pseudoscalar meson is described by a dashed line, a vector meson by a solid line. The cross denotes an NLO vertex, the dot an LO vertex.}
	\label{fig:Diag-for-SE}
	\end{center}
\end{figure}

At chiral order $Q^4$, the self energy for a pseudoscalar meson is given by tree-level diagrams with a $Q^4$-vertex and loop-diagrams with one or two $Q^2$-vertices (Fig.\ \ref{fig:Diag-for-SE}).
%
%\afterpage{
%}
%
In pure $\chpt$, only vertices with even numbers of pseudoscalar mesons are possible for the self energy of a pseudoscalar meson. Therefore, the first pure-$\chpt$ diagram in Fig.\ \ref{fig:Diag-for-SE} is generated by $\Phi^2$-contributions in the NLO Lagrangian $\Lch^{\te{NLO}}$ and the second one, the ``tadpole" diagram, by $\Phi^4$-con\-tri\-bu\-tions in the LO Lagrangian $\Lch^{\te{LO}}$,
\begin{align}
	\Lch^{\te{NLO}, 2\Phi} =& \  \frac{4 B_0}{F_r^2} \left\{ L_4 \bra \mathcal{M} \ket \bra \partial_\mu \Phi \partial^\mu \Phi \ket + L_5 \bra \mathcal{M} \partial_\mu \Phi \partial^\mu \Phi \ket \vphantom{\bra \mathcal{M} \Phi \ket^2} \! \right\} \adb \nn \\
	& - \frac{8 B_0^2}{F_r^2} \left\{ 2\, L_6 \bra \mathcal{M} \ket \bra \mathcal{M} \Phi^2 \ket + 2 \, L_7 \bra \mathcal{M} \Phi \ket^2 \right. \nn \\[-0.75em]
	& \phantom{ - \frac{8 B_0^2}{F^2} \left\{  \right\} } \left. +\,  L_8 \bra \mathcal{M} \Phi \mathcal{M} \Phi + \mathcal{M}^2 \Phi^2 \ket \vphantom{\bra \mathcal{M} \Phi \ket^2} \! \right\} ,\adb \nn \\
	\Lch^{\te{LO},4\Phi} =& \frac{1}{24 F_r^4} \left\{ F_r^2 \bra \left[ \Phi, \partial_\mu \Phi \right] \Phi \partial^\mu \Phi \ket + F^2 B_0 \bra \mathcal{M} \Phi^4 \ket \right\}.
\end{align}
Hereby, all external sources except masses are set to zero such that $\chi {=} \chi^\dagger {=} 2 B_0 \mathcal{M}$. The pure $\chpt$ self energy is calculated in \cite{Scherer:2002tk} and can be expressed as
\begin{align}
	&\SelfEn_\chpt^{(P)}(p^2) =  A(P) + p^2 B(P) \in \order(Q^4)\,, \nn \\
	& B(P) := \sigma(P) + \rho(P) \label{eq:chpt-coeff}
\end{align}
for a given pseudoscalar meson $P$. The formulae for $A(P)$, $\sigma(P)$ and $\rho(P)$ are given as
\begin{align}
	&A(\pi) = \frac{\mpi^2}{F_r^2} \Big\{  - \frac{1}{6} ( \mu_\pi + 2 \mu_K + \mu_\eta ) + 16 (2 \mk^2 + \mpi^2 ) L_6^r \nn \\[-0.7em]
	& \phantom{A(\pi) = \frac{\mpi^2}{F}} \ + 16 \mpi^2 L_8^r \Big\}, \adb \nn \\
	&A(K) =\frac{\mk^2}{F_r^2} \Big\{  \frac{1}{12} ( -3 \mu_\pi - 6 \mu_K + \mu_\eta ) + 16 (2 \mk^2 + 
\mpi^2 ) L_6^r  \adb \nn \\[-0.7em]	
	&  \phantom{A(K) = \frac{\mk^2}{F}} \  + 16 \mk^2 L_8^r \Big\}, \adb \nn \\
	& A(\eta) = -\frac{2 \meta^2 \mu_\eta}{3F_r^2}  + \frac{\mpi^2}{6 F_r^2} (-3 \mu_\pi + 2 \mu_K + \mu_\eta ) \nn \\
	& \phantom{A(\eta) =} \  + \frac{16 \meta^4}{F_r^2} L_8^r + \frac{16 \meta^2}{F_r^2} (2 \mk^2 + \mpi^2) L_6^r  \nn \\ 
	& \phantom{A(\eta) =} \  + \frac{128 (\mk^2 - \mpi^2)}{9 F_r^2} (3 L_7^r + L_8^r) \, , \adb \nn \\[1em]
	&\sigma(\pi) = \frac{1}{3 F_r^2} ( 2\mu_\pi + \mu_K ) \,, \adb \nn \\
	&\sigma(K) = \frac{1}{4 F_r^2} ( \mu_\pi + 2 \mu_K + \mu_\eta) \,, \adb \nn \\
	&\sigma(\eta) = \frac{\mu_K}{F_r^2} \, , \adb \nn \\[1em]
	&\rho(P) = - \frac{8}{F_r^2} [(2 \mk^2 + \mpi^2) L_4^r + \mP^2 L_5^r ]\,, \adb \nn \\[1em]
	&\mu_P := \frac{1}{16\pi^2} \mP^2 \log \left(\frac{\mP^2}{\mu^2}\right) . \label{eq:Def-A-sigma-rho}
\end{align}

For the three diagrams in Fig.\ \ref{fig:Diag-for-SE} including loops with vector mesons, vertices proportional to $V^2 \Phi^2$, $V^2\Phi$ and $V \Phi^2$, respectively, are needed. Expanding $\Lvec$ in orders of $\Phi$ produces no vertex proportional to $V^2 \Phi$, \ie the second diagram does not exist in the framework used in this article. The vertex proportional to $V^2\Phi^2$ includes the commutator $\left[V^{\mu\nu}, \partial^\rho V_{\rho\nu} \right]$ which is equal to zero when the two vector-meson fields are Wick contracted. Thus, the tadpole diagram does not contribute either. Therefore, the one-loop contribution with vector mesons to the self energy of pseudoscalar mesons is only given by the last diagram in Fig.\ \ref{fig:Diag-for-SE} and generated by the Lagrangian
\begin{align}
	\Lvec^{V\! \Phi^2} = i \, \frac{f_V h_P}{8 \,F_r^2} \bra V^{\mu\nu} \partial_\mu \Phi \, \partial_\nu \Phi \ket . \label{eq:Lvec-Vphi2}
\end{align}
For calculating the self energy, the matrix elements \cite{Gasser:1983yg, Ecker:1988te}
\begin{align*}
	&\bra 0 | T V^{\mu\nu}_a(x) V^{\alpha\beta}_a(y) |0 \ket \nn \\
	& \hspace{2em} = - \, \frac{i}{\mv^2} \int \frac{\te{d}^4q}{(2\pi)^4}\, \frac{e^{-iq(x-y)}}{ q^2- \mv^2 } \left\{ (\mv^2- q^2) g^{\mu\alpha} g^{\nu\beta} \right. \nn \\
	& \phantom{\hspace{2em} = } \left. \hspace{5em}  + \, q^\mu q^\alpha g^{\nu\beta} - q^{\mu}q^{\beta} g^{\nu\alpha} - (\mu \leftrightarrow \nu) \right\}, \adb \nn \\
	& \partial^\mu_x \partial^\nu_y \bra 0 | T \phi_P(x) \phi_P(y) |0 \ket = i \int \frac{\te{d}^4q}{(2\pi)^4} \, \frac{e^{-iq(x-y)}}{ q^2 - \mP^2} \, q^\mu q^\nu\,
\end{align*}
for a vector meson and a pseudoscalar meson $P$, respectively, are used. Therewith, the contribution of loops with vector mesons to the self energy of a pseudoscalar meson $P$ reads as
\begin{align}
	\SelfEn_{\te{vec}}^{(P)}(p^2) =& \, -\frac{f_V^2 h_P^2}{16 \pi^2 \cdot 64 F_r^4} \left\{ 4 \delta_{P\pi} \left[ 2 g_\pi(p^2) + g_K(p^2) \right] \right. \nn \\
	& \hspace{0.7em} + \,  3 \delta_{P K} \left[ g_\pi(p^2) + 2g_K(p^2) + g_\eta(p^2) \right] \nn \\
	& \hspace{0.7em} \left. + \, 12 \delta_{P\eta} g_K(p^2) \right\}. \label{eq:SE-vec-full}
\end{align}
The function $g_R$ depends on both the squared momentum $p^2$ of the incoming pseudoscalar meson $P$ and the mass of the pseudoscalar meson $R$ in the loop \cite{Pascual:1984zb, intTab}, 
\begin{align*}
	g_R(p^2) & := 4 \alpha^2_R(p^2) \left[ 1 - \frac{L_R(p^2)}{p^2} \right] \nn \\ 
	& + p^2 ( 3 S_R - p^2 ) \log\frac{\mv \mR}{\mu^2} \nn \\
	& +  D_R \left(3 S_R - \frac{D_R^2}{p^2} \right) \log\frac{\mv}{\vphantom{\mR^2}\mR}
\end{align*}
including the abbreviations 
\begin{align*}
	&L_R(p^2) : = \alpha^{\phantom{2}}_R(p^2) \log \left( \frac{p^2-S_R +2 \alpha^{\phantom{2}}_R(p^2)}{p^2 - S_R - 2\alpha^{\phantom{2}}_R(p^2)} \right), \adb \nn \\
	&\alpha^2_R(p^2) := \frac{1}{4} \left( (\mv - \mR)^2 - p^2 \right) \left( (\mv + \mR)^2 - p^2 \right), \adb \nn \\
	& S_R := \mv^2 + \mR^2 \, , \ \ D_R := \mv^2 - \mR^2 .
\end{align*}

For further calculations, the expansion of the self energy $\SelfEn_{\te{vec}}^{(P)}$ up to (including) chiral order $Q^4$ is of interest. Thereby, the self energy has to be evaluated at $p^2 {=} M_P^2$ for determining the full mass $M_P$. Due to the softness of pseudoscalar mesons, it can be expanded at $p^2 = 0\,$\footnote{Note that $g_R$ is in fact finite at $p^2{=}0$ such that this expansion is possible.}. Since only the contributions in $p^2$ and $p^4$ are non-zero at $\order(Q^4)$, the approximated self energy $\SelfEn_{\te{appr}}^{(P)}$ of loops with vector mesons is given by
\begin{align}
	&\SelfEn_{\te{appr}}^{(P)}(p^2) = [b_0 +b_V(P)] p^2 + c_V p^4 + \order(Q^6) \, , \adb \nn \\
	& b_0 := \frac{6 \beta \, \mv^2 }{16\pi^2 \, F_r^4} \left(1 + 6 \log\frac{\mv^2}{\mu^2} \right) \in \order(1) \,, \adb \nn \\
	& b_V(P) := \frac{\beta}{16\pi^2 \, F_r^4} \left(1 + 6 \log\frac{\mv^2}{\mu^2} \right) \left\{ 2 \delta_{P\pi} (2 \mpi^2 + \mk^2)  \right.  \nn \\
	& \phantom{b_V(P) :=  }\ \  \left. +\,  \delta_{P K} (\mpi^2+ 5 \mk^2) + 6 \delta_{P \eta} \mk^2 \right\} \in \order(Q^2)\,, \adb \nn \\
	& c_V := -\frac{2 \beta}{16 \pi^2 \, F_r^4} \left( 5 + 6 \log\frac{\mv^2}{\mu^2} \right) \in \order(1)\, , \adb \nn \\
	&\beta := -\frac{f_V^2 h_P^2}{64} \label{eq:vec-coeff}
\end{align}
at $\order(Q^4)$. Hereby, the Gell-Mann-Okubo relation \cite{Gasser:1984gg},  $3 \lM_{\! \eta}^2 = 4 \mk^2 + \mpi^2$, was used. Note that in contrast to a pure $\chpt$-calculation the LO mass is not equal to the bare mass but
\begin{align}
	M_P^2 = \frac{\mP^2}{1-b_0} + \order(Q^4)
\end{align} 
whereby the bare mass $\mP^2$ differs from the bare $\chpt$ mass by a factor of $F^2/F_r^2$ (cf. Eq.\ \eqref{eq:Def-bare-mass}). 

\subsection*{Test of renormalisation-point invariance at $\order(Q^4)$}
The pseudoscalar masses calculated with the mass equation \eqref{eq:mass-eq} depend on the chosen renormalisation scale $\mu$ both directly via chiral logarithms $\log( \te{mass}^2 / \mu^2)$ in the contributions to the self energy (cf.\ Eq.\ \eqref{eq:chpt-coeff} and \eqref{eq:SE-vec-full}) and indirectly via the scale dependence of the renormalised low-energy constants $F_r^2$ and $L_i$. Since the mass is a physical observable, it has to be independent of the scale $\mu$, \ie 
\begin{align}
	\frac{\text{d} M_P^2}{ \text{d}\mu} \stackrel{!}{=} 0 \,.
\end{align}
This invariance can be used to verify the underlying theoretical assumptions and to check the calculations carried out so far. The calculations to test $\mu$ independence are performed at $\order(Q^4)$ and in one-loop accuracy. The latter can be most easily traced using the large-$N_c$ counting where $N_c$ denotes the number of colours \cite{Kaiser:2000gs, inf},
\begin{align*}
	F^2, \, f_V^2 \in \order(N_c) \, , \ \ \mv, \mP \in \order(1).
\end{align*}
In large-$N_c$ counting, the low-energy constants can be expanded as
\begin{align}
	&F_r^2 = F^2 + \frac{\varphi \, \mv^2}{F^2} \bar{\lambda} + \order(1/N_c) \,, \adb \nn \\
	&L_i^r = L_i + \left(\frac{1}{2} \Gamma_i + \frac{\Lambda_i}{F^2} \right) \bar{\lambda} +\order(1/N_c) \, , \adb \nn \\
	& \varphi , \, \Lambda_i \in \order(N_c)\, .
\end{align}
Therewith, the dependence on a scale $\mu$ for a given low-energy constant $c_r = c_0 + \gamma \bar{\lambda}$ reads as
\begin{align*}
c_r(\mu) = \left[c_r(\mu_0) + \frac{2 \gamma}{16\pi^2} \log(\mu_0) \right] - \frac{2\gamma}{16\pi^2} \log(\mu)
\end{align*}
for an arbitrary reference scale $\mu_0$. $\gamma$ can be re{\-}con{\-}struct{\-}ed from Eq.\ \eqref{eq:Ren-F} and \eqref{eq:Ren-const}.

The mass equation for a pseudoscalar meson $P$ can be solved analytically at $\order(Q^4)$, \ie using the approximated vector-loop contribution $\SelfEn_{\te{appr}}^{(P)}$ to the self energy instead of the full contribution $\SelfEn_{\te{vec}}^{(P)}$. At $\order(Q^4)$, the mass is given as
\begin{align*}
	M_P^2 =&\ \mP^2 + A(P) + \left[ B(P) + b_0 + b_V(P) \right] \mP^2 + c_V \mP^4 \nn \\
	&+ \order(Q^6) + \order(1/N_c^2).
\end{align*}
Note that the second possible solution of the quadratic equation for $M_P^2$ has a non-zero contribution at $\order(1)$ and is therefore not considered here. 

For $\mu$ invariance, the derivative of $M_P^2$ with respect to $\mu$ has to vanish at each order separately, \ie both at $\order(Q^2)$ and at $\order(Q^4)$. The resulting equations can be reformulated into relations for the renormalisation parameters (cf.\ Eq. \eqref{eq:Ren-F} and \eqref{eq:Ren-const}) and the coefficient $\beta$ of the approximated self energy with vector-meson loops (cf.\ Eq.\ \eqref{eq:vec-coeff}), 
\begin{align}
	\varphi = 36 \beta + \order(N_c^0) \, , \ 4 \Lambda_4 = 3 \beta + \order(N_c^0) 
\end{align}
whereby the first relation describes renormalisation-point invariance at $\order(Q^2)$ and the second the additional condition at $\order(Q^4)$. Hereby, the correlation
\begin{align*}
	\Lambda_4 = \frac{1}{3} \Lambda_5 = -3 \Lambda_7 = \Lambda_8
\end{align*}
was used \cite{inf}. The relations above are fulfilled by the values for $\varphi$, $\beta$ and $\Lambda_4$ as given in Eq.\ \eqref{eq:Ren-F}, Eq.\ \eqref{eq:vec-coeff} and Tab.\ \eqref{tab:chpt-le-constants}, respectively. Thus, the calculated mass $M_P^2$ is renormalisation-point invariant at $\order(Q^4)$ as necessary for a physical observable. 

\section{Contributions to decay constants of pseudoscalar mesons} \label{sec:decay-const}
\begin{figure}[h]
 \begin{center} $
 \begin{array}{ccc}
 \includegraphics[trim = 0 22.5 0 -22.5, width=0.12\textwidth]{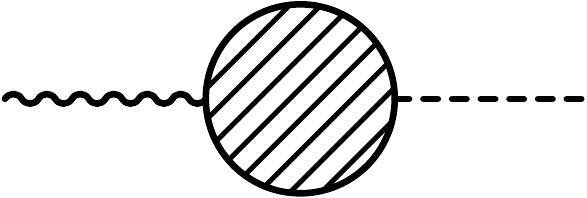} \vspace{1em} &
\large{\ = \ } & 
\includegraphics[trim = 0 22.5 0 -22.5,  width=0.12\textwidth]{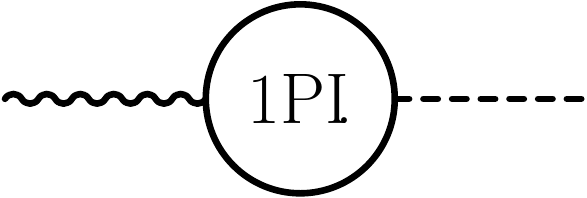}  \hspace{0.5em} \large{\ \times \ } \hspace{0.5em}  \includegraphics[trim = 0 22.5 0 -22.5,  width=0.12\textwidth]{prop-allg.pdf}
 \end{array} $
%[trim = l b r t]
 \caption{General diagram for calculating decay constants. The wiggled line denotes a weak field $a_\mu$, ``1PI" one-particle irreducible diagrams.}
 \label{fig:dc-allg}
 \end{center}
\end{figure}

In this section, the decay constants of pseudoscalar mesons are calculated in general, the numerical results are discussed in section \ref{sec:Numerics}. Decay constants of pseudoscalar mesons can be calculated by Feynman diagrams with an incoming weak field $a_\mu$ and an outgoing pseudoscalar meson as shown on the left-hand side in Fig.\ \ref{fig:dc-allg}. As illustrated on the right-hand side in Fig.\ \ref{fig:dc-allg}, this general diagram can be split into the product of one-particle irreducible (1PI) diagrams with an incoming weak field and an outgoing meson and a meson propagator, \ie a diagram with an incoming and outgoing meson as considered in the previous section.

Denoting the results from the 1PI diagrams with an incoming weak field and an outgoing pseudoscalar field in Fig.\ \ref{fig:dc-allg} as $S_\mu$, the full matrix element for calculating the decay constant $\hat{F}(P)$ of a pseudoscalar meson $P$ reads as
\begin{align}
	i \mathcal{M}_\mu^{(P)} = \frac{i S_\mu^{(P)}(p^2)}{p^2 - M_P^2 + i0^+ } =: \frac{ i p_\mu \hat{F}(P)}{p^2- M_P^2 + i0^+} . \label{eq:ME-dc-allg}
\end{align}
Hereby, $p^2$ denotes the squared momentum of in- and outgoing fields. The physical mass $M_P$ used in the definition of $\hat{F}$ has been determined in the previous section.

\begin{figure}[!h]
\vspace{1em}
	\begin{center} $
\hspace{-1em}	
	\begin{array}{llccc}
		\text{(I)} & \text{pure } \chpt \text{:} \vspace{1em} &
		\hspace{-1.5em} \includegraphics[trim = 0 0 0 0, width=0.08\textwidth]{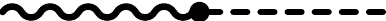} &
		\hspace{-5.5em} \includegraphics[trim = 0 3 0 -3, width=0.08\textwidth]{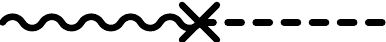} &
		\hspace{-3.5em} \includegraphics[trim = 0 1 0 -1, width=0.08\textwidth]{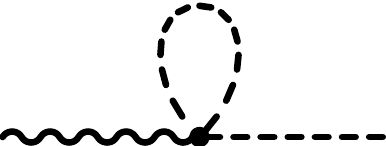}	\\
		\text{(II)} & \multicolumn{4}{l}{\text{loops with vector mesons:}} \\[0.5em] 
		& \hspace{0.5em} \multirow{2}{*}{\includegraphics[trim = 0 -1 0 1, width=0.08\textwidth]{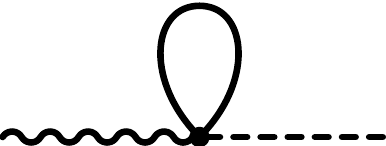}} &
		\hspace{0.5em} \multirow{2}{*}{\includegraphics[trim = 0 11 0 -11, width=0.12\textwidth]{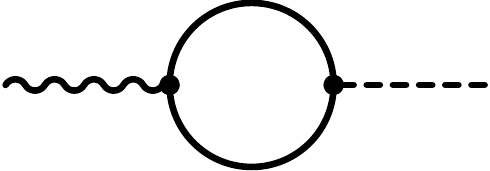}} &
		\hspace{0.5em} \multirow{2}{*}{\includegraphics[trim = 0 11 0 -11, width=0.12\textwidth]{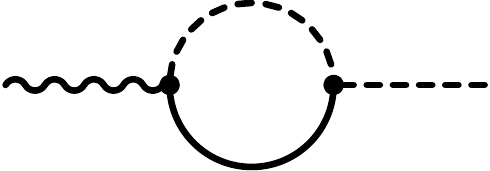}}  & \\[2em]		
	\end{array} $
% width = x, x, 3/2x, 3/2x	
	\caption{Contributing diagrams to $S_\mu^{(P)}$ as defined in Eq.\ \eqref{eq:ME-dc-allg} from pure $\chpt$ (I) and from loops including vector mesons (II), respectively.  A pseudoscalar meson is described by a dashed line, a vector meson by a solid line and the weak field $a_\mu$ by a wiggled line. The dot denotes an LO vertex, the cross an NLO vertex.}
	\label{fig:Diag-for-dc}
	\end{center}
\end{figure}

In Fig.\ \ref{fig:Diag-for-dc}, the one-loop diagrams contributing to $S_\mu$ are listed up to $\order(Q^4)$. For the pure-$\chpt$ diagrams, the necessary parts of the Lagrangian are given by
\begin{align*}
	\Lch^{\te{LO}, a\Phi} &= - F_r  \bra a_\mu \partial^\mu \Phi \ket , \\
	\Lch^{\te{NLO}, a\Phi} &= - \frac{4 \, B_0 F^2}{F_r^3} \left[ 2 (2m + m_s) L_4 \bra a_\mu \partial^\mu \Phi \ket \right. \\[-1em]
	&\phantom{= - \frac{4 \, B_0 F_0^2}{F^3} \left[  \right.} \left. + L_5 \bra \mathcal{M} \left\{ a_\mu, \partial^\mu \Phi \right\} \ket \right], \\
	\Lch^{\te{LO},a\Phi^3} &= \frac{1}{12^3 F_r} \bra a_\mu \left[ \left\{ \Phi^2, \partial^\mu \Phi \right\} - 2 \Phi ( \partial^\mu \Phi) \Phi \right] \ket.
\end{align*}
The first diagram involving vector mesons in Fig.\ \ref{fig:Diag-for-dc} requires an $aV^2 \Phi$-vertex, the second one an $aV^2$- and a $V^2 \Phi$-vertex, and the third one an $aV\Phi$- and a $V\Phi^2$-vertex. As for the mass calculation, the diagram with a vector-meson tadpole is equal to zero because the vertex $aV^2\Phi$ is proportional to the vanishing commutator $\left[V^{\mu\nu}, \partial^\rho V_{\rho\nu} \right]$. Since an $aV^2$-vertex does not exist in $\Lvec$, the second diagram does not contribute either. For the third diagram, the necessary vertices are generated by 
\begin{align*}
	&\Lvec^{V\! \Phi^2} = i \, \frac{f_V h_P}{8 \,F_r^2} \bra V^{\mu\nu} \partial_\mu \Phi \, \partial_\nu \Phi \ket, \nn \\
	&\Lvec^{a \! V \! \Phi} = - i \,\frac{f_V h_P}{4 F_r} \bra \left[ a_\mu, \partial_\nu \Phi \right] V^{\mu\nu} \ket + i \frac{f_V}{2 F_r} \bra \left[ \partial_\mu a_\nu, \Phi \right] V^{\mu\nu} \ket.
\end{align*}
Note that the second term in $\Lvec^{a \! V \! \Phi}$ yields zero in all calculations.

$S_\mu^{(P)}$ is calculated using renormalised perturbation theory \cite{Peskin:1995ev} where the field $\Phi$ is replaced by the renormalised field $\hat{\Phi}$ via
\begin{align}
	&\Phi \mapsto \hat{\Phi} := \sum_P \frac{\lambda^P \phi_P}{\sqrt{Z(P)}}, \adb \nn \\
	&\sqrt{2} \, \lambda_{\pi^+} := -\left( \lambda_1 + i \lambda_2 \right), \, \sqrt{2} \, \lambda_{\pi^-} := \left( \lambda_1 - i \lambda_2 \right),\adb \nn \\
	&\sqrt{2} \, \lambda_{K^+} := -\left( \lambda_4 + i \lambda_5 \right), \, \sqrt{2} \, \lambda_{K^-} := \left( \lambda_4 - i \lambda_5 \right), \adb \nn \\
	&\sqrt{2} \, \lambda_{K^0} := -\left( \lambda_6 + i \lambda_7 \right), \, \sqrt{2} \, \lambda_{\bar{K}^0} := -\left( \lambda_6 - i \lambda_7 \right), \adb \nn \\
	&\lambda_{\pi^0} := \lambda_3, \, \lambda_\eta := \lambda_8
\end{align}
with the Gell-Mann matrices $\lambda_1, \ldots, \lambda_8$. Hereby, the wave-function-renormalisation constant $Z(P)$ is defined via the propagator as \cite{Scherer:2002tk}
\begin{align*}
	& i \Delta(p^2) =: \frac{i Z_P}{p^2-M_P^2} + \left( \te{terms regular at $p^2 = M_P^2$} \right), \\
	& Z_P^{-1} = 1 - \SelfEn'(M_P^2) = 1 - B(P) - \SelfEn_{\te{vec}}^{(P)}{}\! ' (M_P^2)
\end{align*}
with $B {=} \sigma {+} \rho$ for $\sigma$ and $\rho$ given in Eq.\ \eqref{eq:Def-A-sigma-rho} and $\SelfEn_{\te{vec}}^{(P)}$ as defined in Eq.\ \eqref{eq:SE-vec-full}. $\SelfEn'$ denotes the derivative with respect to the squared momentum $p^2$. The Lagrangians needed to calculate the one-loop contributions to $S_\mu$ can now be rewritten in terms of the renormalised field $\hat{\Phi}$. Thereby, the LO-$\chpt$ contribution reads as
\begin{align}
	& \Lch^{\te{LO}, a\Phi} = - F_r \sum_P \sqrt{Z(P)} \, a^\mu_P \,\partial_\mu \hat{\Phi}_P. \label{eq:Def-deltaF}
\end{align}
All remaining terms in the Lagrangians needed for calculating $S_\mu$ have to be rewritten in terms of the renormalised field as well. In principle, their parameters are also multiplied with factors of $\sqrt{Z(P)}$. However, all these terms are already one-loop or NLO contributions. Therefore, the constant $Z(P)$ can be approximated by one for these terms. Therewith, $S_\mu^{(P)}$ for a pseudoscalar meson $P$ can be determined as
\begin{align}
	S_\mu^{(P)}(p^2) = & i p_\mu \Big\{ F_r \sqrt{Z(P)} - F_r \left[ 2 \sigma(P) + \rho(P) \right]  \nn \\[-0.5em]
	& \phantom{ i p_\mu \left\{ \right.} - F_r \, \frac{\SelfEn_{\te{vec}}^{(P)}(p^2)}{p^2} \Big\}. \label{eq:Res-S-afterRen}
\end{align}
Furthermore, the LSZ-reduction formula \cite{Lehmann:1954rq} is applied to the matrix element given in Eq.\ \eqref{eq:ME-dc-allg} yielding
\begin{align*}
	\lim_{p^2 \mapsto M_P^2} \left\{ \left(p^2 - M_P^2 \right) \cdot i\mathcal{M}_\mu^{(P)} \right\} =  i p_\mu \hat{F}(P) = i S_\mu^{(P)}(M_P^2).
\end{align*}
Therewith, the decay constant $\hat{F}(P)$ of a pseudoscalar meson $P$ in one-loop approximation including loops with vector mesons can be determined as 
\begin{align}
	\frac{\hat{F}(P)}{F_r} = 1 & + \frac{1}{2} \SelfEn_{\te{vec}}^{(P)}{}\!'(M_P^2) - \frac{\SelfEn_{\te{vec}}^{(P)}(M_P^2)}{M_P^2} \nn \\
	&- \frac{1}{2} \left[ 3 \sigma(P)+\rho (P) \right] . \label{eq:dc-result}
\end{align}

\section{Numerical results for masses and decay constants} \label{sec:Numerics}

Within this section, the numerical values for both masses and decay constants of pseudoscalar mesons are discussed. In particular, the dependence of masses and decay constants on the bare pion mass $\mpi$ are examined. According to Eq.\ \eqref{eq:Def-bare-mass} this is equivalent to a variation in the light quark mass $m$. This is exactly the situation which is of interest for lattice-QCD calculations -- except for lattice artefacts like finite-volume effects that we do not address in the present work. All calculations within this section are done at a fixed renormalisation point $\mu {=} \mv$. 

As discussed in section \ref{sec:Intro}, two numerical comparisons will be performed in this article: First, differences between static and dynamical vector mesons are considered (subsection \ref{subsec:Num-NNLO}). Second, a pure $\chpt$ calculation at order $Q^4$ is compared to a calculation including loops with vector mesons (subsection \ref{subsec:Num-NLO}). This allows to examine how quantitatively important higher order contributions from vector-meson loops are for formal $Q^4$ calculations. Both comparisons are addressed as functions of the bare pion mass, \ie as functions of the averaged up- and down-quark mass. 

\subsection{Static versus dynamical vector mesons} \label{subsec:Num-NNLO}
\begin{figure}[t]
	\centering
	\includegraphics[trim = 0 35 0 0, width=0.14\textwidth]{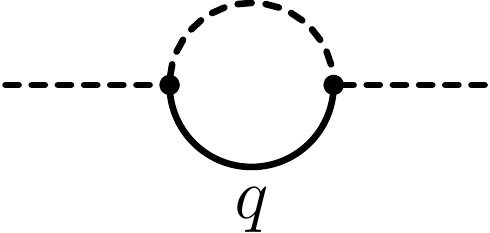} \hspace{0.7em}
	\Large{${\Rightarrow}$} \hspace{0.25em} 
	\includegraphics[trim = 0 12.25 0 -15, width=0.16\textwidth]{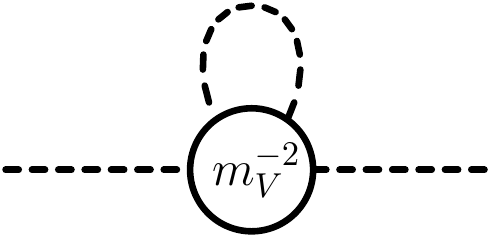} \\[2em]
	\hspace{-9.5em} \includegraphics[trim = 0 -80 0 80, width=0.16\textwidth]{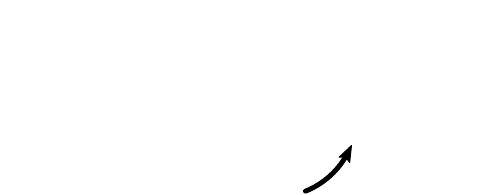} \\[-3em]
	\caption{Transformation of a one-loop diagram with a vector meson to a pure $\chpt$ diagram for $q^2 \ll \mv^2$. Here, $q$ denotes the momentum of the vector meson in the loop on the left-hand side. Note that the circle with the label $\mv^{-2}$ should not be misinterpreted as a vector-meson loop. This circle represents a vertex.}
	\label{fig:loop-gr-q2}
\end{figure}

To compare differences of calculations with static and dynamical vector mesons, static vector mesons are discussed first. In pure $\chpt$ and in the resonance saturation picture \cite{Ecker:1988te, Donoghue:1988ed}, the masses of the pseudoscalar mesons are assumed to be much smaller than the masses of the vector mesons, $M_P \ll \mv$. Furthermore, all involved momenta have to be much smaller than the vector-meson mass as well, $q^2 \ll \mv^2$. Thus, the vector-meson propagator $(\mv^2 - q^2)^{-1}$ can be approximated by $1/\mv^2$. Using this approximation, the one-loop diagram with a vector meson shown at the left-hand side in Fig.\ \ref{fig:loop-gr-q2}  has the same form as a pure $\chpt$-tadpole diagram. Recall from sections \ref{sec:masses} and \ref{sec:decay-const} that the diagram depicted on the left-hand side in Fig.\ \ref{fig:loop-gr-q2} is the only one-loop diagram with vector mesons contributing to masses and decay constants of pseudoscalar mesons. For $q^2 {\ll} \mv^2$ and using the modified minimal subtraction scheme of \cite{Gasser:1984gg} we provide the finite contribution to the pseudoscalar self energy coming from the vector-meson-loop diagram with the vector-meson propagator shrank to a point. It is given by
\begin{align}
	& \SelfEn^{(P)}_{\te{point}}(p^2) = \frac{f_V^2 h_P^2}{16 \pi^2 \cdot 128 F_r^4} \frac{p^2}{\mv^2} \left\{ 4 \delta_{P\pi} [2 h(\pi) + h(K)] \right. \nn \\ 
	& \phantom{M^{\te{approx}}_P =} \left. \, + \, 3 \delta_{PK} [h(\pi) + 2 h(K) + h(\eta)] + 12 \delta_{P\eta} \, h(K) \right\} \! , \adb \nn \\
	& h(R) := \mR^4 \left( 1 + 6 \log \frac{\mR^2}{\mu^2} \right).	\label{eq:SE-point}
\end{align}

In a pure $\chpt$ calculation, all degrees of freedom except the low-lying pseudoscalar mesons are integrated out, \ie their contributions are included in the low-energy constants. Thus, in the resonance saturation picture the numerical value of the vector-meson-loop diagram with a point-like propagator shown on the right-hand side in Fig.\ \ref{fig:loop-gr-q2} has to be the same as the corresponding contributions from pure $\chpt$ at N$^2$LO. These corresponding contributions are pseudoscalar tadpole diagrams with a vertex of $\order(Q^4)$ and have already been calculated in \cite{Amoros:1999dp} for both pseudoscalar masses and decay constants. They were used as cross checks for the calculations done within the present article. Thereby, the values for the non-vanishing low-energy constants in the resonance saturation picture which are needed within this article are given by \cite{Ecker:1988te}\footnote{Note that the constant $G_V$ used in \cite{Ecker:1988te} is equal to $\frac{1}{4} f_V h_P$.}
\begin{align}
	L_1^{V} = \frac{f_V^2 h_P^2}{128 \mv^2}, \ L_2^V = 2 L_1^V, \ L_3^V = -6 L_1^V.  \label{eq:const-rs}
\end{align}

For dynamical vector mesons, the full propagator $(q^2 - \mv^2)^{-1}$ is used for calculating the loop diagram shown on the left-hand side in Fig.\ \ref{fig:loop-gr-q2}. Both loop diagrams with an approximated and with a full propagator will depend on the parameters $h_P$, $f_V$, $F_r$, $\mv$ and the bare pseudoscalar masses $\mpi$ and $\mk$. For the qualitative comparison between the two types of diagrams, the corresponding calculations are normalised such that they do not depend on $h_P$, $f_V$ and $F_r$ anymore.  

The differences between loop diagrams with approximated and full propagators are considered as functions of the bare pion mass $\mpi$ in comparison to the reference point $\mpi = M_\pi^{\te{exp}}$. For that let $T_P$ denote the normalised contribution from the tadpole diagram with the shrunk propagator to the mass or decay constant of a pseudoscalar meson $P$ and $I_P$ the normalised contribution from the full vector-meson-loop diagram. For masses, the contributions of the corresponding diagrams to the pseudscalar self energy are considered in $T_P$ and $I_P$, \ie
\begin{align*}
	&T_P(\mpi^2) := \frac{16 \pi^2 \cdot 128 F_r^4}{f_V^2 h_P^2} \, \SelfEn_{\te{point}}^{(P)}(M_P^2), \adb \\
	&I_P(\mpi^2) := \frac{16 \pi^2 \cdot 128 F_r^4}{f_V^2 h_P^2} \, \SelfEn_{\te{vec}}^{(P)}(M_P^2)
\end{align*}
with $\SelfEn_{\te{point}}$ and $\SelfEn_{\te{vec}}$ as defined in Eq.\ \eqref{eq:SE-point} and \eqref{eq:SE-vec-full}, respectively. For decay constants, the contributions are given by (cf.\ Eq.\ \eqref{eq:dc-result})
\begin{align*}
	&T_P(\mpi^2) := \frac{16 \pi^2 \cdot 128 F_r^4}{f_V^2 h_P^2} \! \left[ \frac{1}{2} \SelfEn^{(P)}_{\te{point}}{}\!\!\!\!'\,(M_P^2) - \frac{\SelfEn_{\te{point}}^{(P)}(M_P^2)}{M_P^2} \right] \! , \adb \\
	&I_P(\mpi^2) := \frac{16 \pi^2 \cdot 128 F_r^4}{f_V^2 h_P^2} \left[ \frac{1}{2} \SelfEn^{(P)}_{\te{vec}}{}\!'(M_P^2) - \frac{\SelfEn_{\te{vec}}^{(P)}(M_P^2)}{M_P^2} \right] \! .
\end{align*}
Hereby, $\SelfEn^{(P)}_{\te{point/vec}}{}\hspace{-2.2em}' \hspace{1.8em}$ denotes the derivative with respect to the squared momentum $p^2$. For $M_P$ we use the respective bare mass $\mP$.
The differences between the contributions from the diagrams with approximated and with full propagators compared to the reference point $\mpi = M_\pi^{\te{exp}}$ can now be expressed via the functions
\begin{align*}
	&\Delta T_P \big(\mpi^2 \big) := T_P \big(\mpi^2 \big) - T_P \big(\mpi^2 = (M_\pi^{\te{exp}})^2 \big), \adb \\
	&\Delta I_P \big(\mpi^2 \big) := I_P \big(\mpi^2 \big) - I_P \big(\mpi^2 = (M_\pi^{\te{exp}})^2 \big).
\end{align*}

\begin{figure}[!b]
	\centering
	\includegraphics[width = 0.48\textwidth, trim = 60 70 40 50]{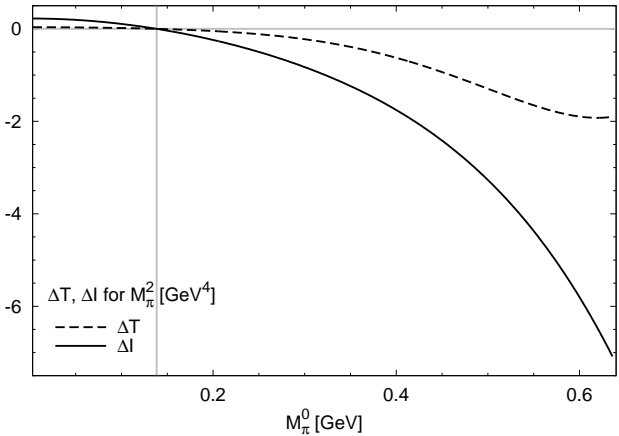}
	\caption{$\Delta T_\pi$ and $\Delta I_\pi$ for the squared pion mass as a function of the bare pion mass. The vertical line represents the experimental pion mass $M_\pi^{\te{exp}} = 138 \, \te{MeV}$ which is taken as the reference point.}	
	\label{fig:mass-pi-fctMpi}
	%Fig.6
\end{figure}

\begin{figure}[!t]
	\centering
	\includegraphics[width = 0.48\textwidth, trim = 60 70 40 50]{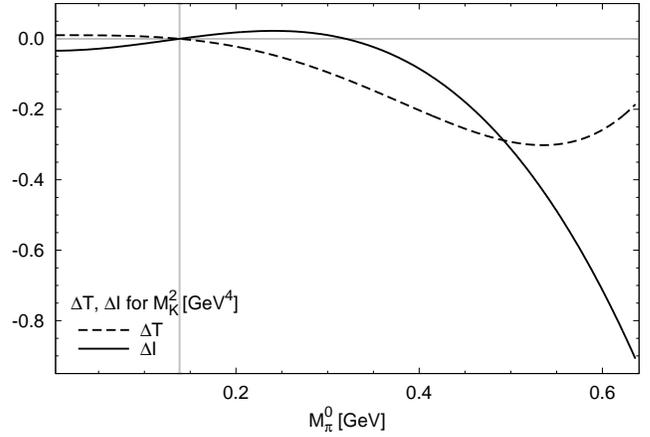}
	\caption{Same as in Fig.\ \ref{fig:mass-pi-fctMpi} but for the squared kaon mass.}	
	\label{fig:mass-kaon-fctMpi}
	%Fig.7
\end{figure}

\begin{figure}[!t]
	\centering
	\includegraphics[width = 0.48\textwidth, trim = 60 70 40 44]{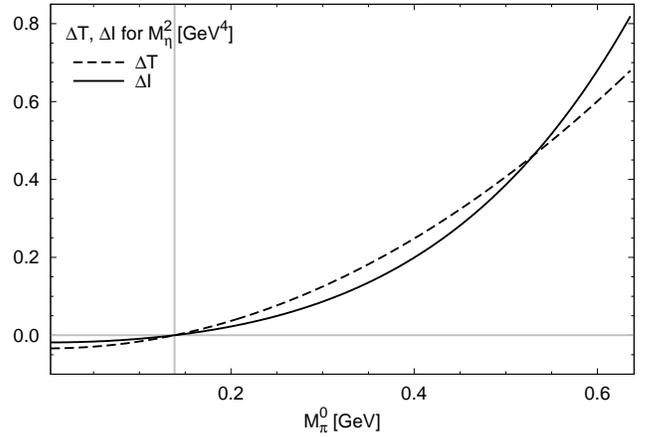}
	\caption{Same as in Fig.\ \ref{fig:mass-pi-fctMpi} but for the squared mass of the $\eta$-meson.}	
	\label{fig:mass-eta-fctMpi}
	%Fig.8
\end{figure}

\begin{figure}[!h]
	\centering
	\includegraphics[width = 0.48\textwidth, trim = 60 70 40 44]{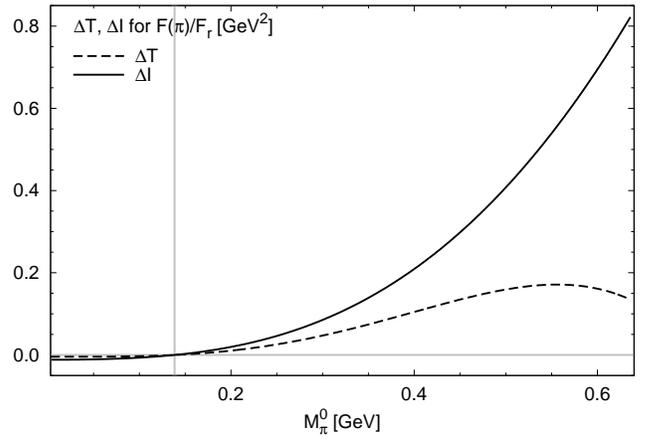}
	\caption{Same as in Fig.\ \ref{fig:mass-pi-fctMpi} but for the pion decay constant $\hat{F}(\pi)$}		
	\label{fig:dc-pi-fctMpi}
	%Fig.9
\end{figure}

\begin{figure}[!b]
	\centering
	\includegraphics[width = 0.48\textwidth, trim = 60 70 40 50]{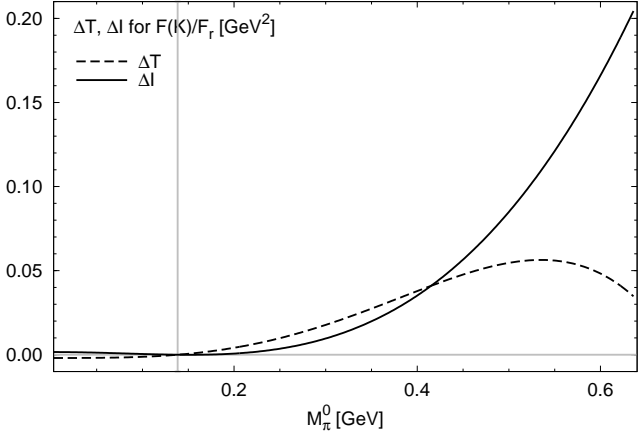}
	\caption{Same as in Fig.\ \ref{fig:mass-pi-fctMpi} but for the kaon decay constant $\hat{F}(K)$.}	
	\label{fig:dc-kaon-fctMpi}
	%Fig.10
\end{figure}

To calculate the dependence of $\Delta T$ and $\Delta I$ on $\mpi$ only, the bare kaon mass $\mk$ as the remaining free parameter is chosen to be equal to the physical kaon mass $M_K^{\te{exp}} {=} 496\, \te{MeV}$.
In Fig.\ \ref{fig:mass-pi-fctMpi} - \ref{fig:dc-kaon-fctMpi}, $\Delta T$ and $\Delta I$ are plotted for pseudoscalar masses and decay constants of pions, kaons, and $\eta$-mesons. For the $\eta$-meson, only a kaon is possible in the vector-meson-loop diagram contributing to the self energy. Therefore, the decay constant of the $\eta$-meson depends only on the kaon mass and, thus, $\Delta T_\eta(\mpi^2) {=} \Delta I_\eta(\mpi^2) {=} 0$. The mass of the $\eta$-meson depends in addition on the bare $\eta$-mass and therewith on the pion mass. For all observables, there are differences between the calculation with the approximated propagator ($\Delta T$) and the full calculation ($\Delta I$) significant for pion masses above approximately $250 \, \te{MeV}$. 

\begin{figure}[!b]
	\centering
	\includegraphics[width = 0.48\textwidth, trim = 60 70 40 50]{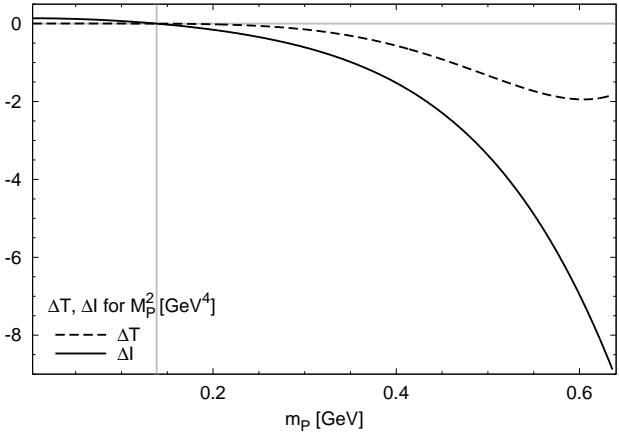}
	\caption{$\Delta T_P$ and $\Delta I_P$ for the squared pseudoscalar mass $M_P^2$ as a function of the bare mass $m_P$ with reference point $M_\pi^{\te{exp}}$. The vertical line represents the experimental pion mass $M_\pi^{\te{exp}} = 138 \, \te{MeV}$. }	
	\label{fig:mass-fctMp-MpiExp}
	%Fig.11
\end{figure}

\begin{figure}[!h]
	\centering
	\includegraphics[width = 0.48\textwidth, trim = 60 70 40 50]{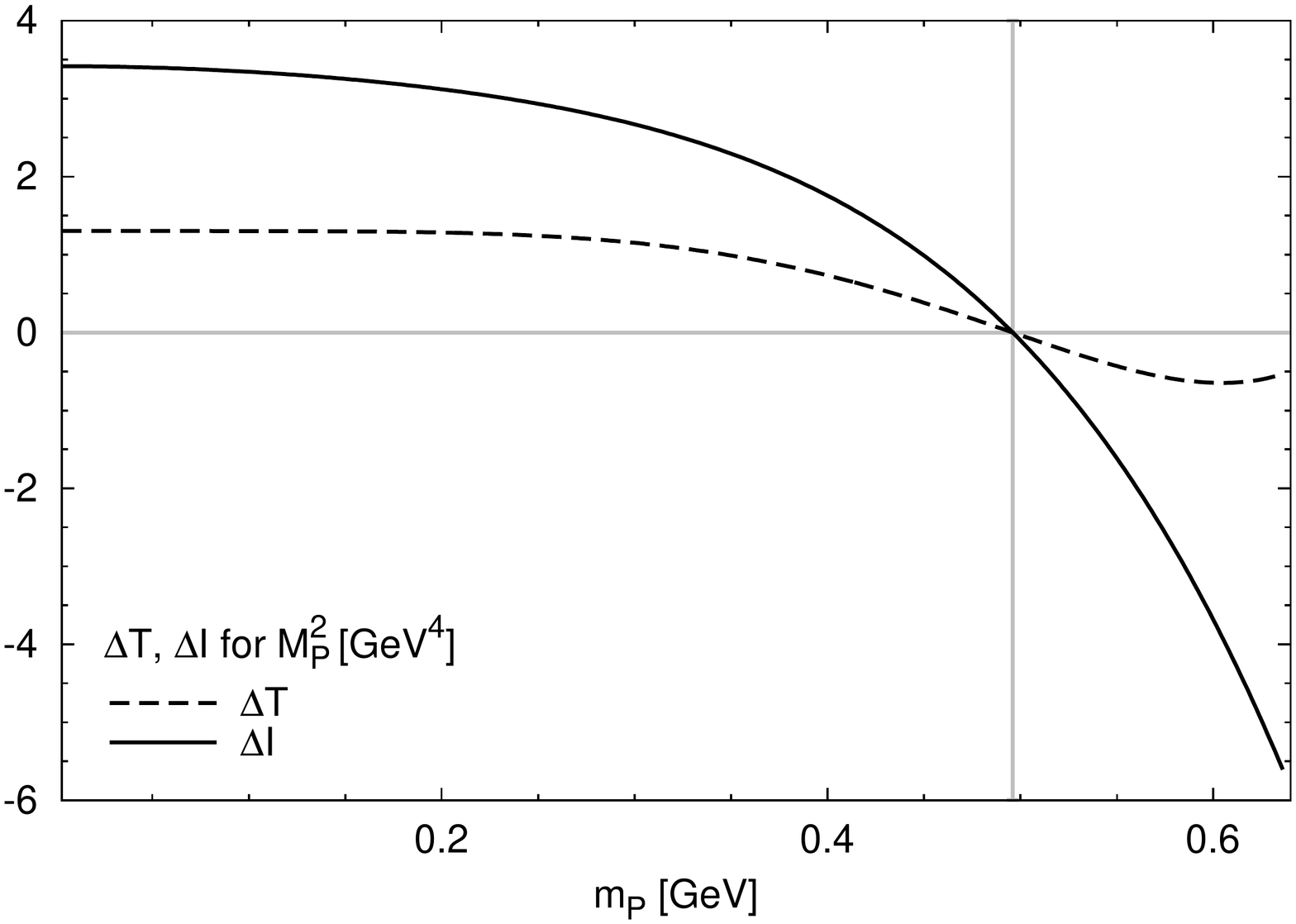}
	\caption{Same as in Fig.\ \ref{fig:mass-fctMp-MpiExp} but for the reference point $M_K^{\te{exp}} = 496 \, \te{MeV}$ represented by the vertical line.
	%The vertical line represents the experimental kaon mass $M_K^{\te{exp}} = 496 \, \te{MeV}$.
	}	
	\label{fig:mass-fctMp-MkExp}
	%Fig.12
\end{figure}

\begin{figure}[!h]
	\centering
	\includegraphics[width = 0.48\textwidth, trim = 60 70 40 60]{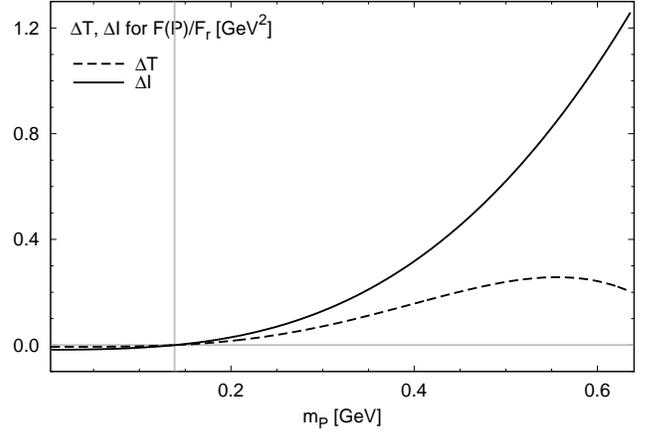}
	\caption{Same as in Fig.\ \ref{fig:mass-fctMp-MpiExp} bur for the pseudoscalar decay constant $\hat{F}(P)$ for the reference point $M_\pi^{\te{exp}}$. The vertical line represents the experimental pion mass $M_\pi^{\te{exp}} = 138 \, \te{MeV}$.}	
	\label{fig:dc-fctMp-MpiExp}
	%Fig.13
\end{figure}

\begin{figure}[!h]
	\centering
	\includegraphics[width = 0.48\textwidth, trim = 60 70 40 69.5]{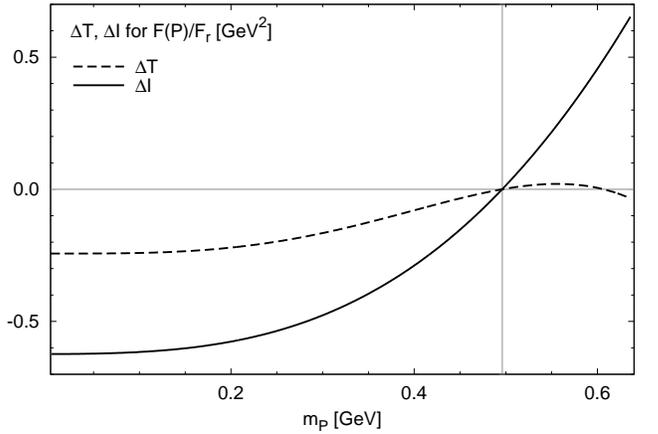}
	\caption{Same as in Fig.\ \ref{fig:dc-fctMp-MpiExp} but for the reference point $M_K^{\te{exp}} = 496 \, \te{MeV}$ represented by the vertical line.}	
	\label{fig:dc-fctMp-MkExp}
	%Fig.14
\end{figure}

\indent
Additionally, we study the differences between $\Delta T$ and $\Delta I$ in the SU(3)-symmetric case, \ie for $\mpi {=} \mk {=:} m_P$. Thereby, the masses and decay constants of all three pseudoscalar mesons become the same. As a reference point, either $m_P {=} M_\pi^{\te{exp}}$ or $m_P {=} M_K^{\te{exp}}$ is chosen. In Fig.\ \ref{fig:mass-fctMp-MpiExp} - \ref{fig:dc-fctMp-MkExp}, $\Delta T$ and $\Delta I$ for the SU(3)-symmetric mass and decay constant are plotted for the two different reference points. While for the reference point $M_\pi^{\te{exp}}$ all results are small for pion masses smaller than approximately $250 \, \te{MeV}$, they will already be visible for small pion masses if the reference point $M_K^{\te{exp}}$ is taken.
\subsection{Comparison between pure NLO-$\chpt$ calculations and calculations with vector-meson loops} \label{subsec:Num-NLO}

While in the previous subsection the differences between one-loop diagrams with static and dynamical vector mesons are discussed, the diagram with dynamical vector mesons, \ie with the full vector-meson propagator will be compared in the following to an NLO-$\chpt$ tadpole diagram with an LO vertex. Again, the dependence of both diagrams on the bare pion mass $\mpi$ is studied. 

\newpage
Assuming that calculations are carried out in \textit{the} effective field theory for both pseudoscalar and vector mesons, the difference between a scenario with and without vector mesons at a given order can only be visible in different values of the low-energy constants. Therefore, the low-energy constants at chiral orders $Q^2$ and $Q^4$ can be adjusted such that the two scenarios yield the same results for observables up to (including) order $Q^4$.

In general, a physical quantity at chiral order $Q^4$ can be expressed as
\begin{align*}
	\te{quantity} &= \# F_r^2(\mu) + \# L_i^r(\mu) + (\te{$\chpt$ loops})(\mu) \\
	& \phantom{=} \, + (\te{vector loops})(\mu) \adb \\
	&= \# \big( F_0^2 + \delta F^2(\mu) \big) + \# \big( L_i^{\chpt}(\mu) + \delta L_i(\mu) \big) \nn \\
	& \phantom{=} \, + (\te{$\chpt$ loops})(\mu) + (\te{vector loops})(\mu)	
\end{align*}
depending on the renormalisation scale $\mu$. The low-energy constants $F_0^2$ and $L_i^{\chpt}$ are those defined by pure $\chpt$. Thus, the deviations $\delta F^2$ and $\delta L_i$ from these constants can be adjusted such that they cancel the contributions from the vector loops once the vector loops are expanded in chiral orders up to $\order(Q^4)$,
\begin{align}
	\te{quantity} &= \# F_0^2 + \# L_i^{\chpt}(\mu) + (\te{$\chpt$ loops})(\mu) \nn \\
	& \phantom{=} \, + \left\{ (\te{vector loops})(\mu) \right. \nn \\
	& \phantom{=} \, \phantom{+ \{} \left. - (\te{vector loops at }\order(Q^4))(\mu) \right\} . \nn   
\end{align}
The difference between the approximated contribution from vector loops and the full contribution has to be at least of chiral order $Q^6$. It can be explored how quantitatively important differences of order $Q^6$ are for a formal $Q^4$ calculation if pure-$\chpt$ calculations and calculations with vector mesons are compared. The aim of this subsection is to perform such a comparison for the masses and decay constants of pseudoscalar mesons. Hereby, the mass of a pseudscalar meson $P$ is given as the solution of the mass equation
\begin{align}	
	0 = M_P^2 - \left[ \mP^2 + \SelfEn_{\chpt}^{(P)}(M_P^2) + \SelfEn_{\te{vec}}^{(P)}(M_P^2) - \SelfEn_{\te{appr}}^{(P)}(M_P^2) \right] \label{eq:mass-eq-NLO-comp}
\end{align}
with the $\chpt$ self energy $\SelfEn_{\chpt}$ defined in Eq.\ \eqref{eq:chpt-coeff}, the full vector-meson self energy $\SelfEn_{\te{vec}}$ defined in Eq.\ \eqref{eq:SE-vec-full} and the approximated vector-meson self energy at $\order(Q^4)$, $\SelfEn_{\te{appr}}$, defined in Eq.\ \eqref{eq:vec-coeff}. The decay constant of a pseudoscalar meson $P$ is given by
\begin{align}
	\frac{\hat{F}(P)}{F_0} =& \, 1 + \frac{1}{2} \left[ \SelfEn_{\te{vec}}^{(P)}{}\!'(M_P^2) - 2 M_P^{-2} \SelfEn_{\te{vec}}^{(P)}(M_P^2) \right] \nn \\
	& \phantom{1} \, - \frac{1}{2} \left[ \SelfEn_{\te{appr}}^{(P)}{}\!\!\!'(M_P^2) - 2 M_P^{-2} \SelfEn_{\te{appr}}^{(P)}(M_P^2) \right] \nn \\
	& \phantom{1} \, - \frac{1}{2} \left[ 3 \sigma(P) + \rho(P) \right] \label{eq:F0-NL0-comp}
\end{align}
including the contributions $\sigma(P)$ and $\rho(P)$ to the $\chpt$ self energy (\ref{eq:Def-A-sigma-rho}) with the $L_i^r$ replaced by $L_i^{\chpt}$. The contribution including the approximated vector-meson self energy can be expressed as
\begin{align*}
	\SelfEn_{\te{appr}}^{(P)}{}\!\!\!'(M_P^2) - 2 \, \frac{\SelfEn_{\te{appr}}^{(P)}(M_P^2)}{M_P^2} = - \left( b_0 + b_V(P) \right) + \order(Q^4)
\end{align*}
with the functions $b_0$ and $b_V$ as defined in \eqref{eq:vec-coeff}.

Both the equations for mass and decay constant depend on the parameters
\begin{align*}
	F_0^2, \, \mpi, \, \mk, \, L_i^{\chpt}, \, h_P, \, f_V.
\end{align*}
For practical matters, $\mu {=} \mv$ and the standard values for $L_i^{\chpt}$ (cf.\ Tab.\ \ref{tab:chpt-le-constants}) are chosen. The two parameters $f_V$ and $h_P$ can either be determined by comparison with experimental data yielding $f_V {=} 150 \, \te{MeV}$ and $h_P {=} 1.50$ \cite{Lutz:2008km}\footnote{Note that the parameter $h_P$ was redefined compared to the definition used in \cite{Lutz:2008km}.} or by using the KSFR relation yielding $f_V {\approx} \sqrt{2} F^{\te{exp}}_\pi$ and $h_P {=} 2$ \cite{Ecker:1989yg}. As in the previous subsection, the bare pion mass $\mpi$ is taken as a running parameter. In principle, both the bare pion and the bare kaon mass will differ from the pure $\chpt$ value by a factor of $F^2/F_r^2$ (cf.\ \eqref{eq:Def-bare-mass}) if loops with vector mesons are taken into account. Here, however, a one-loop approximation is considered and all changes in the low-energy constants of $\chpt$ are assumed to be cancelled by the vector-loop contributions up to $\order(Q^4)$. Therefore, the bare masses are equal to their $\chpt$ result. 

We still have to decide how to choose our remaining parameters $F_0$ and $\mk$ when $\mpi$ is varied. In principle, it would be appealing to readjust $F_0$ and $\mk$ such that specific observables remain constant. For instant, one might consider to keep mass and decay constant of the kaon at their physical values. We have found, however, that this leads to numerically rather unstable results. In subsection \ref{subsec:det-F0} we will elaborate on these problems, yet for a somewhat simplified case. We study the SU(3) symmetric case and keep the pseudoscalar decay constant at the experimental value for the kaon decay constant. After facing all the problems related to this choice, we will discuss an alternative, namely we decide to keep $F_0$ and $B_0 m_s$ constant when varying $\mpi$. This is discussed in subsection \ref{subsec:NLO-results} below. Clearly neither $F_0$ nor $B_0 m_s$ are observables, which puts a grain of salt in our analysis.

\subsubsection{Determining $F_0$ in the SU(3)-symmetric case $\mpi = \mk$} \label{subsec:det-F0}

\begin{figure}[!b]
	\centering
	\includegraphics[width = 0.48\textwidth, trim = 60 70 40 80]{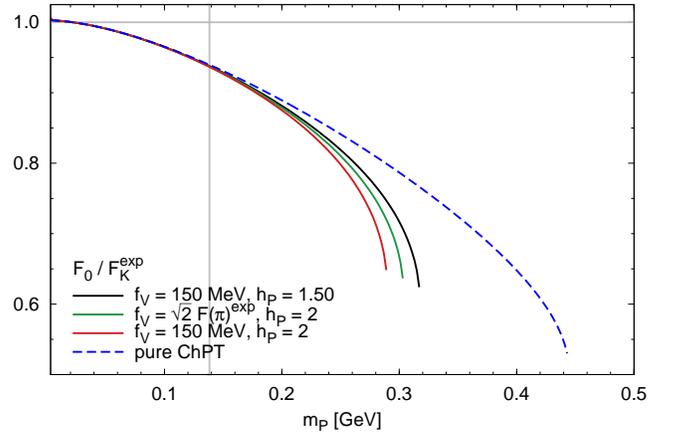}
	\caption{$F_0$ as a function of the bare pseudoscalar mass $m_P$ for both a pure $\chpt$ calculation (blue dashed line) and calculations with vector mesons (solid lines). The different colours for the calculation with vector mesons represent the results for the different values for $f_V$ and $h_P$ whereby they are in the same order as in the legend. The vertical line represents the experimental pion mass $M_\pi^{\te{exp}} = 138 \, \te{MeV}$.}	
	\label{fig:NLO-f0}
	%Fig.15
\end{figure}

As discussed before, the parameter $F_0$ is determined as a function of $m_P := \mpi = \mk$ by assuming that the experimental value $F_K^{\te{exp}} = 110 \, \te{MeV}$ for the kaon decay constant is reproduced exactly. The decay constant is given as (cf.\ Eq.\ \eqref{eq:F0-NL0-comp})
\begin{align}
	\frac{\hat{F}(P)}{F_0} =& \, 1 + \frac{1}{2} \left[ \SelfEn_{\te{vec}}^{(P)}{}\!'(M_P^2) - 2 M_P^{-2} \SelfEn_{\te{vec}}^{(P)}(M_P^2) \right] \nn \\
	& \phantom{1} \, - \frac{1}{2} \left[ \SelfEn_{\te{appr}}^{(P)}{}\!\!\!'(M_P^2) - 2 M_P^{-2} \SelfEn_{\te{appr}}^{(P)}(M_P^2) \right] \nn \\
	& \phantom{1} \, - \frac{1}{2} \left[ 3 \sigma(P) + \rho(P) \right] \nn \\
	=&\!: \,  1 + F_0^{-2} G + F_0^{-4} H.   \label{eq:Def-H-G}
\end{align}
Hereby, $F_0^{-2} G$ denotes the contribution from pure $\chpt$ and $F_0^{-4} H$ the one from loops with vector mesons. Recall from section \ref{sec:masses} that the self energy for loops with vector mesons is proportional to $F_0^{-4}$ and the functions $\sigma(P)$ and $\rho(P)$ defined in Eq.\ \eqref{eq:Def-A-sigma-rho} are proportional to $F_0^{-2}$. Therefore, both $G$ and $H$ are independent of $F_0$. Note further that neither of these functions depends on the chosen pseudoscalar meson $P$ in the SU(3)-symmetric case. It turns out that the result for $F_0$ in a pure $\chpt$ calculation has no real solution for values of $m_P$ between approximately $440 \, \te{MeV}$ and $750 \, \te{MeV}$ (cf.\ Fig.\ \ref{fig:NLO-f0}). For a calculation including vector mesons the result for $F_0$ has a non-vanishing imaginary part already for values of $m_P$ larger than approximately $280 \, \te{MeV}$ (cf.\ Fig.\ \ref{fig:NLO-f0}). Thereby, all calculations depend on the chosen values for $f_V$ and $h_P$.

For illustration, consider the case of pure $\chpt$. The result for $F_0$ is then given by\footnote{Only the solution for $F_0$ is used which is equal to $F_K^{\te{exp}}$ in the limit of vanishing one-loop corrections.}
\begin{align}
	F_0 = \frac{1}{2} \left( F_K^{\te{exp}} + \sqrt{ (F_K^{\te{exp}})^2 - 4 G} \, \right). \label{eq:F0-result}
\end{align}
Since both $\sigma(P)$ for $m_P < \mv$ and $\rho(P)$ are negative, $G$ is positive for $m_P < \mv$. Therefore, the square root and therewith the result for $F_0$ can become imaginary. It is interesting to see that even for (SU(3)-symmetric) pure $\chpt$ this happens already before $m_P$ reaches the physical kaon mass $M_K^{\te{exp}} {=} 496 \, \te{MeV}$. 

\begin{figure}[!b]
	\centering
	\includegraphics[width = 0.48\textwidth, trim = 60 70 40 50]{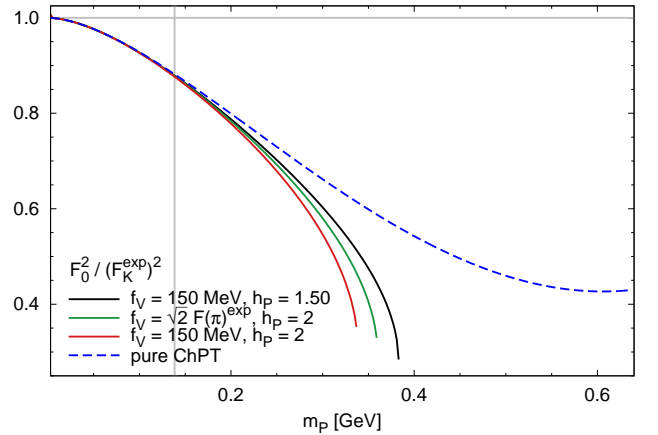}
	\caption{Same as in Fig.\ \ref{fig:NLO-f0} but for the squared parameter $F_0^2$ determined with Eq.\ \eqref{eq:F02-withSqr}.}	
	\label{fig:NLO-f02-withSqr}
	%Fig.16
\end{figure}

\begin{figure}[!b]
	\centering
	\includegraphics[width = 0.48\textwidth, trim = 60 70 40 50]{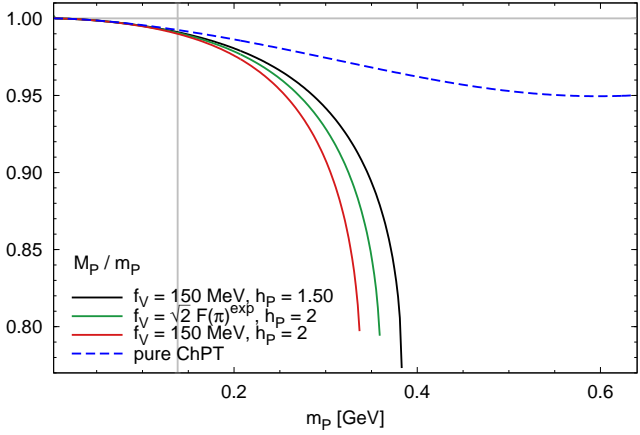}
	\caption{Mass of a pseudoscalar meson as a function of the bare pseudoscalar mass $m_P$ for both a pure $\chpt$ calculation (blue dashed line) and calculations with vector mesons (solid lines). For determining $F_0^2$, Eq.\ \eqref{eq:F02-withSqr} is used. The different colours for the calculation with vector mesons represent the results for the different values for $f_V$ and $h_P$ whereby they are in the same order as in the legend. The vertical line represents the experimental pion mass $M_\pi^{\te{exp}} = 138 \, \te{MeV}$.}	
	\label{fig:NLO-mass-withSqr}
	%Fig.17
\end{figure}

In \eqref{eq:F0-result}, $F_K^{\te{exp}}$ is a non-vanishing quantity at LO of $\chpt$ while $G$ is an NLO quantity. Therefore, one definitely leaves the regime of applicability of the power counting when $G$ becomes as big as $F_K^{\te{exp}}$. Thus, one might think about some rearrangements. Instead of determining $F_0$ we determine $F_0^2$ in the following. This is appealing in the sense that it is $F_0^2$ and not $F_0$ which appears as a low-energy constant in the Lagrangian of $\chi$PT. Starting from a correlator of quark currents and saturating it with a one-Goldstone-boson state leads also directly to the equation for ${\hat F}^2$, see, e.g., \cite{Gasser:1986vb}. At one-loop accuracy Eq.\ \eqref{eq:Def-H-G} is equivalent to 
\begin{align}
	\hat{F}^2(P) = F_0^2 + 2 G + 2 F_0^{-2} H. \label{eq:quadrateq}
\end{align}
Therewith, $F_0$ as a function of the bare pion mass can be determined as
\begin{align}
	F_0^2 = \frac{1}{2} \left\{ (F_K^{\te{exp}})^2 - 2 G + \sqrt{\left[(F_K^{\te{exp}})^2 - 2G \right]^2 - 8 H} \right\}. \label{eq:F02-withSqr}
\end{align}
The solution for $F_0^2$ is now purely real in pure $\chpt$ while the solution including loops with vector mesons still has a non-vanishing imaginary part for bare pion masses larger than approximately $330 \, \te{MeV}$ (cf.\ Fig.\ \ref{fig:NLO-f02-withSqr}).
\hspace{-0.985em} We also checked whether it is possible that the non-physical parameter $F_0^2$ is complex but the physical observables depending on $F_0^2$ are real. However, the mass of a pseudoscalar meson $P$ will become complex as well for bare pion masses above $330 \, \te{MeV}$ if loops with vector mesons are taken into account and Eq.\ \eqref{eq:F02-withSqr} is used to determine $F_0^2$ (cf.\ Fig.\ \ref{fig:NLO-mass-withSqr}).

\begin{figure}[!b]
	\centering
	\includegraphics[width = 0.48\textwidth, trim = 60 70 40 50]{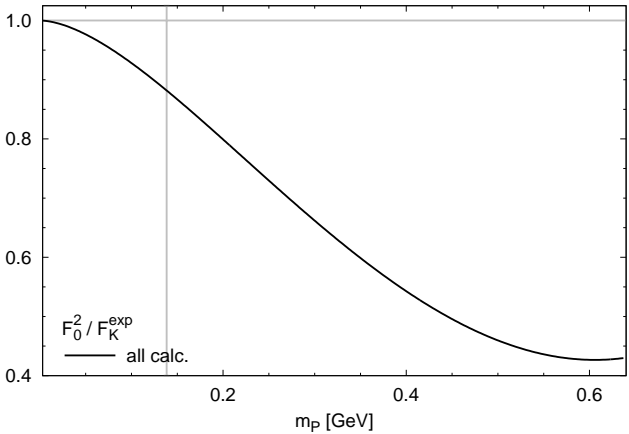}
	\caption{$F_0^2$ determined with Eq.\ \eqref{eq:F02-withoutSqr} as a function of the bare pseudoscalar mass $m_P$. Here, the calculations with and without vector mesons do not differ visibly. The vertical line represents the experimental pion mass $M_{\pi}^{\te{exp}} = 138 \, \te{MeV}$.}	
	\label{fig:NLO-f02-withoutSqr}
	%Fig.18
\end{figure}

\begin{figure}[!b]
	\centering
	\includegraphics[width = 0.48\textwidth, trim = 60 70 40 50]{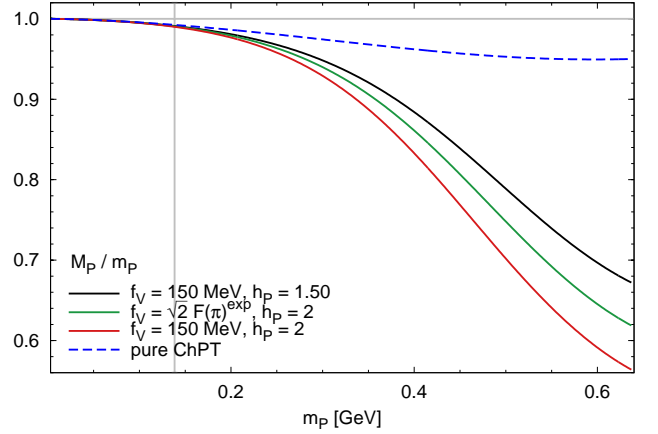}
	\caption{Same as in Fig.\ \ref{fig:NLO-mass-withSqr} but using \eqref{eq:F02-withoutSqr} for determining $F_0^2$.}	
	\label{fig:NLO-mass-withoutSqr}
	%Fig.19
\end{figure}

\begin{figure}[!b]
	\centering
	\includegraphics[width = 0.48\textwidth, trim = 60 70 40 50]{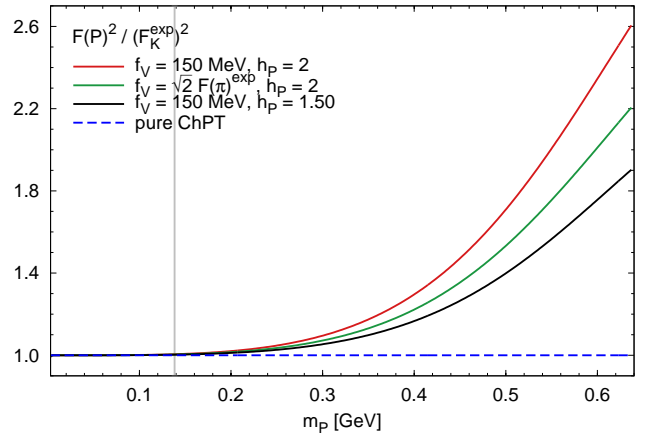}
	\caption{Same as in Fig.\ \ref{fig:NLO-mass-withoutSqr} but for the squared decay constant of a pseudoscalar meson calculated with \eqref{eq:quadrateq}.}	
	\label{fig:NLO-dc-withoutSqr}
	%Fig.20
\end{figure}

To avoid non-vanishing imaginary parts of $F_0^2$ for calculations with vector mesons, the square root in the solution \eqref{eq:F02-withSqr} for $F_0^2$ can be expanded in one-loop accuracy as well. Then, $F_0^2$ simplifies to
\begin{align}
	F_0^2 = (F_K^{\te{exp}})^2 - 2 G - 2 (F_K^{\te{exp}})^{-2} H. \label{eq:F02-withoutSqr}
\end{align}
In this case, all results for $F_0^2$ are real (cf.\ Fig.\ \ref{fig:NLO-f02-withoutSqr}).
This is also reflected in the purely real results for the pseudoscalar mass shown in Fig.\ \ref{fig:NLO-mass-withoutSqr}. Note that the results for the mass differ depending on whether loops with vector mesons are taking into account and, if yes, which values for $f_V$ and $h_P$ are used, while the results for $F_0^2$ do not differ visibly. Although $F_0^2$ determined with \eqref{eq:F02-withoutSqr} is now real for all calculations performed within this article, the solution is not satisfactory, either. Recall that the value for $F_0^2$ is determined by assuming that the calculation for the decay constant resembles the experimental value for the kaon decay constant exactly. Thus, the decay constant calculated with \eqref{eq:quadrateq} should at least approximately yield the experimental kaon decay constant if $F_0^2$ as given in \eqref{eq:F02-withoutSqr} is inserted. However, due to the additional expansion in one-loop accuracy the calculation including loops with vector mesons deviates significantly from the experimental kaon decay constant at large bare pion masses (cf.\ Fig.\ \ref{fig:NLO-dc-withoutSqr}). Note that for a pure $\chpt$ calculation no second expansion is necessary such that the pure-$\chpt$ result resembles the kaon decay constant. 

All in all, it does not seem to be possible to determine $F_0$ or $F_0^2$ for interesting values of the bare pion mass in a self-consistent way if loops with vector meson are included. As discussed before, problems occur already for a pure $\chpt$ calculation. Therefore, a fixed value for $F_0$ is used in the following to calculate masses and decay constants of pseudoscalar mesons.

Note that in Figs.\ \ref{fig:NLO-f0} - \ref{fig:NLO-dc-withoutSqr} one can see that the chiral limit is correctly approached. For $m_P \rightarrow 0$ all curves fall on top of each other. This cannot be observed in the plots in the following part where the strange quark mass is kept fixed.

\subsubsection{Masses and decay constants for $\mpi \neq \mk$} \label{subsec:NLO-results}

\begin{figure}[!t]
	\centering
	\includegraphics[width = 0.48\textwidth, trim = 60 70 40 50]{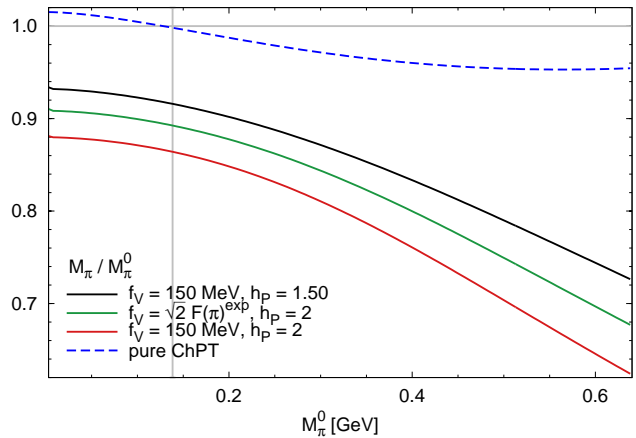}
	\caption{Pion mass as a function of the bare pion mass for both a pure $\chpt$ calculation (blue dashed line) and calculations with vector mesons (solid lines). The different colours for the calculation with vector mesons represent the results for the different values for $f_V$ and $h_P$ whereby they are in the same order as in the legend. The vertical line represents the experimental pion mass $M_\pi^{\te{exp}} = 138 \, \te{MeV}$.}	
	%Fig.21
	\label{fig:NLO-mass-pion}
\end{figure}

\begin{figure}[!t]
	\centering
	\includegraphics[width = 0.48\textwidth, trim = 60 70 40 70]{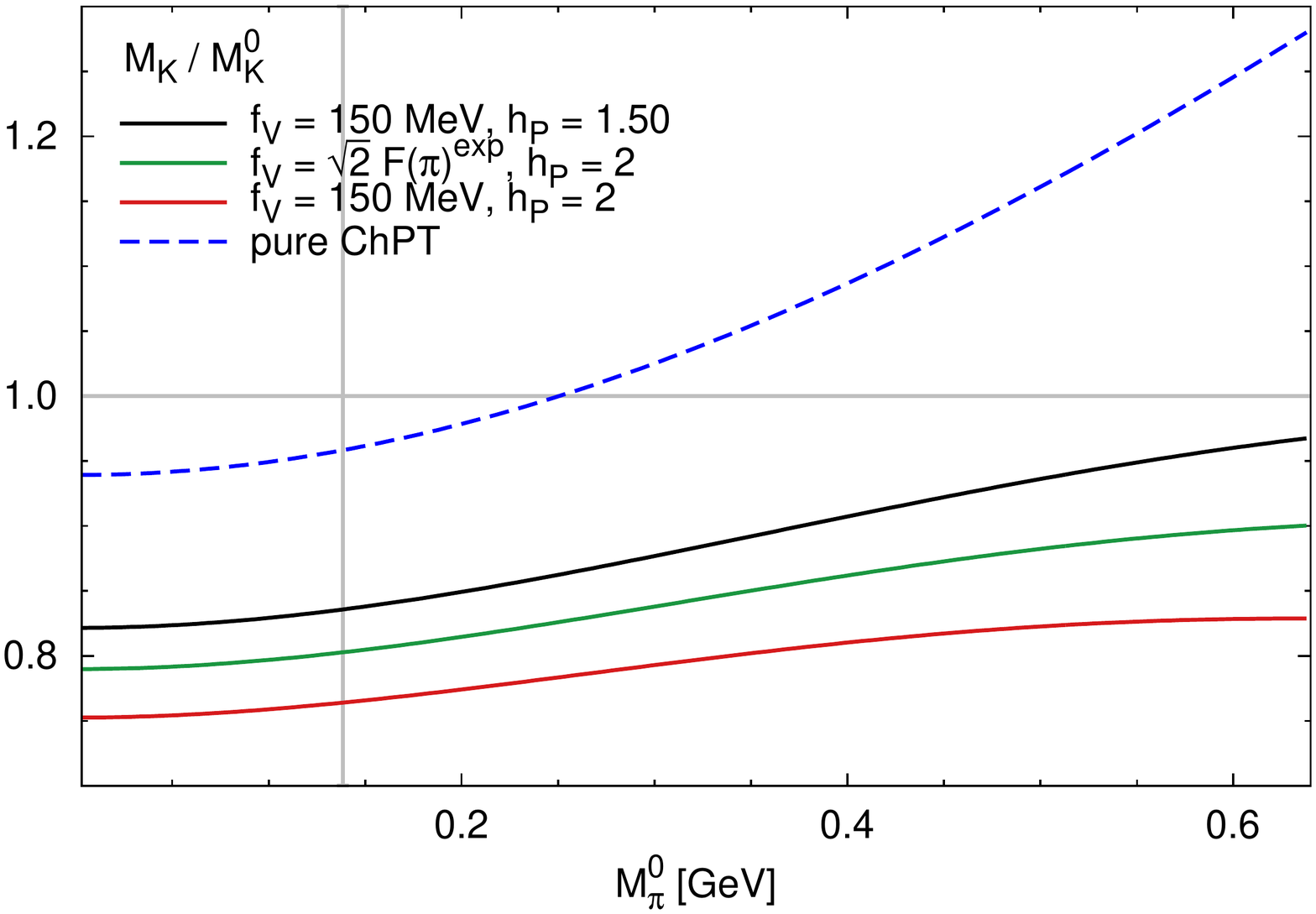}
	\caption{Same as in Fig.\ \ref{fig:NLO-mass-pion} but for the kaon mass.}
	\label{fig:NLO-mass-kaon}
	%Fig.22
\end{figure}

\begin{figure}[H]
	\centering
	\includegraphics[width = 0.48\textwidth, trim = 60 70 40 85]{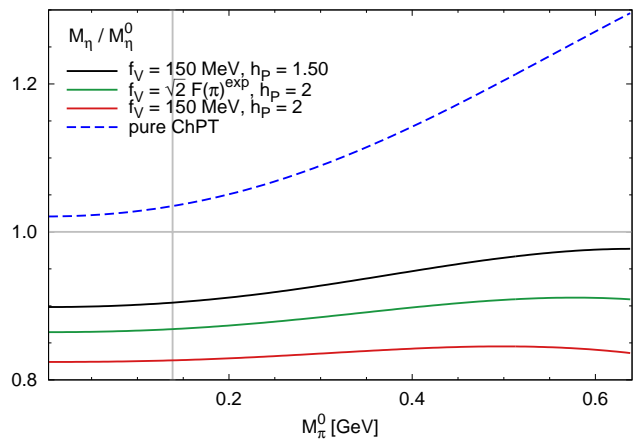}
	\caption{Same as in Fig.\ \ref{fig:NLO-mass-pion} but for the $\eta$-meson mass.}
	\label{fig:NLO-mass-eta}
	%Fig.23
\end{figure}

\begin{figure}[!t]
	\centering
	\includegraphics[width = 0.48\textwidth, trim = 60 70 40 50]{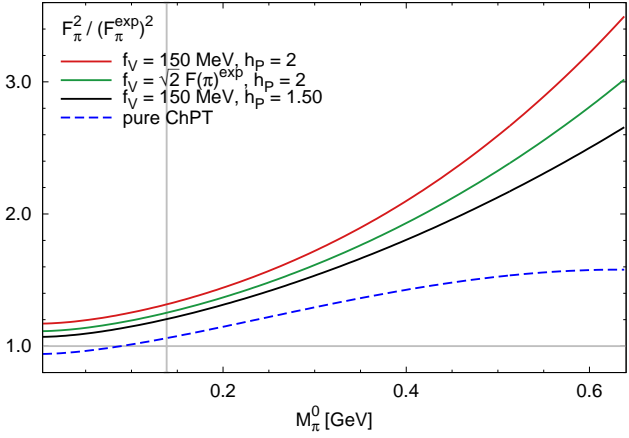}
	\caption{Same as in Fig.\ \ref{fig:NLO-mass-pion} but for the squared pion decay constant normalised to $F_\pi^{\te{exp}} {=} 92 \, \te{MeV}$.}
	\label{fig:NLO-dc-pion}
	%Fig.24
\end{figure}

\begin{figure}[!t]
	\centering
	\includegraphics[width = 0.48\textwidth, trim = 60 70 40 50]{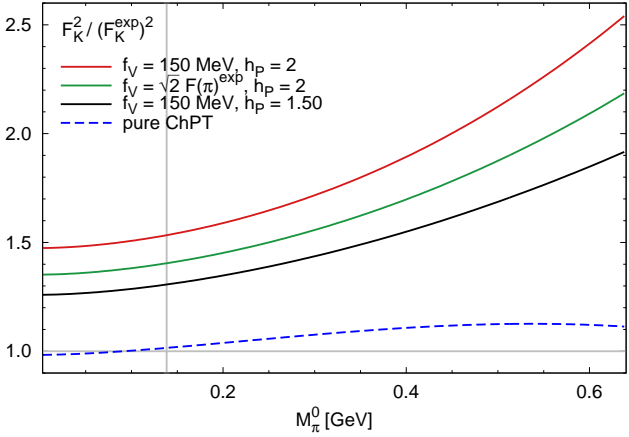}
	\caption{Same as in Fig.\ \ref{fig:NLO-mass-pion} but for the squared kaon decay constant normalised to $F_K^{\te{exp}} {=} 110 \, \te{MeV}$.}
	\label{fig:NLO-dc-kaon}
	%Fig.25
\end{figure}

\begin{figure}[H]
	\centering
	\includegraphics[width = 0.48\textwidth, trim = 60 70 40 50]{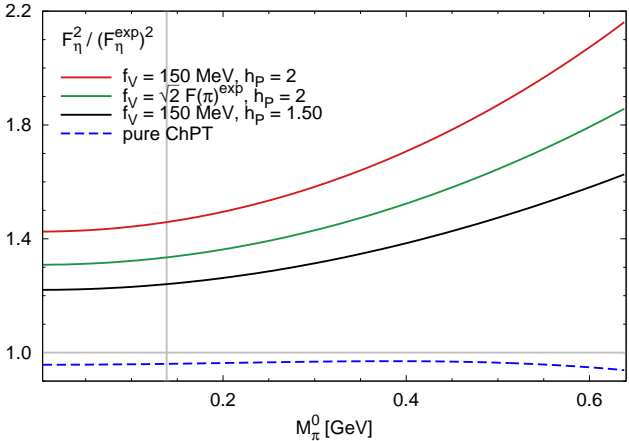}
	\caption{Same as in Fig.\ \ref{fig:NLO-mass-pion} but for the squared $\eta$-meson decay constant. As a reference value, $F_\eta^{\te{exp}} {:=} 1.3 F_\pi^{\te{exp}} {=} 120\, \te{MeV}$ as calculated in \cite{Gasser:1984gg} is used.}
	\label{fig:NLO-dc-eta}
	%Fig.26
\end{figure}

Because of the problems discussed in the previous subsection we now use fixed values for $F_0$ and $B_0 m_s$. Therefore, the value $F_0 {=} 81\, \te{MeV}$ is used \cite{Bijnens:2014lea} in accordance with the values for the low-energy constants $L_i^{\chpt}$ (cf.\ Tab.\ \ref{tab:chpt-le-constants}). To determine $B_0 m_s$ we note that the bare kaon mass can be expressed as
\begin{align}
	\mk^2 = \frac{1}{2} \mpi^2 + B_0 m_s. \label{eq:Def-mk}
\end{align}
Thereby, the mass $m_s$ of the strange quark is determined by the experimental pion and kaon mass \cite{PDG2015},
\begin{align}
	&B_0 m_s = (M_K^{\te{exp}})^2 - \frac{1}{2} (M_\pi^{\te{exp}})^2.
	%, \nn \\
	%&M_K^{\te{exp}} = 496 \, \te{MeV}, \, M_\pi^{\te{exp}} = 138\, \te{MeV}.
	\label{eq:Def-ms}
\end{align}
Then, the mass of a pseudoscalar meson can be determined by solving the mass equation \eqref{eq:mass-eq-NLO-comp} and its squared decay constant can be determined using Eq.\ \eqref{eq:quadrateq}. The masses and squared decay constants as functions of the bare pion mass are shown in Figs.\ \ref{fig:NLO-mass-pion} - \ref{fig:NLO-dc-eta}. The masses are normalised to the $\chpt$-LO results, \ie to the masses $\mP$, and the squared decay constants to the experimental values $(F_P^{\te{exp}})^2$. Thereby, the kaon mass in LO is given in \eqref{eq:Def-mk} as a function of the bare pion mass and the strange quark mass $B_0 m_s$ while the Gell-Mann-Okubo relation is used for the $\eta$ mass at LO (cf.\ \eqref{eq:Def-bare-mass}). The deviation of the pure $\chpt$ calculation from unity shows the difference between an LO and an NLO calculation. The deviation of the pure $\chpt$ calculation from the calculation with vector mesons on the other hand shows a difference which is formally of N$^2$LO. If vector mesons were not important, \ie if the $\chpt$ convergence was good, the difference between the pure $\chpt$ and the vector-meson calculation would be less than the difference between the NLO and the LO calculation, \ie the deviation of the pure $\chpt$ calculation from unity. However, for all quantities this is not the case already at regions with low bare pion masses. This illustrates the importance of including dynamical vector mesons in low-energy calculations. The deviations between the vector-meson calculations for different values of $h_P$ and $f_V$ are always smaller than the deviations from pure $\chpt$. This indicates the robustness of our qualitative finding about the importance of vector mesons. 

\newpage
The first impression from Figs.\ \ref{fig:NLO-mass-pion} - \ref{fig:NLO-dc-eta} concerning the importance of vector mesons might be that either the results are just wrong or that this points to a fundamental flaw of $\chi$PT. This, however, would be a misinterpretation. What seems to be most astonishing is the result for the properties of the pion, Figs.\ \ref{fig:NLO-mass-pion} and \ref{fig:NLO-dc-pion}. Already at low quark masses, \textit{i.e.}, at small values of the bare pion mass, one sees a significant deviation of the pion mass from the bare pion mass. 
On the other hand, the rule of thumb seems to tell that effects should be suppressed by powers of the pion mass over a 
typical hadronic scale. In our case one might use here the vector-meson mass or $4\pi F$. Yet this point of view is somewhat oversimplified. As 
a matter of fact, already NLO $\chi$PT --- for three flavors --- predicts that in the two-flavor chiral limit the deviation of 
the ratio $(M_\pi / \mpi)^2$ from unity is driven by $\mu_{K,\eta}/(4\pi F)^2$, see \eqref{eq:chpt-coeff} and \eqref{eq:Def-A-sigma-rho}.
Obviously this effect survives the two-flavor chiral limit. Besides the logarithm it provides a factor $M_{K,\eta}^2/(4\pi F)^2$. 

Correspondingly we can expect that the loops with vector mesons generate deviations from unity that scale with $Q^4/(m_V^2 \, (4\pi F)^2)$. The power of four in this estimate comes from the fact that we have compensated all vector-meson effects of NLO. The remaining N$^2$LO effect scales with $Q^4$ relative to LO. 
We have also included the typical factor $(4\pi F)^2$ from the loop and added appropriate powers of $m_V$ to make the ratio 
dimensionless. If $Q$ takes the value of the kaon mass, this dimensionless ratio $Q^4/(m_V^2 \, (4\pi F)^2)$ will not be very small. A 10\% \ effect appears rather reasonable and this is what we observe in the two-flavor chiral limit in Fig.\ \ref{fig:NLO-mass-pion}. Replacing in our estimate $Q$ by the kaon mass is induced by the loop contribution of a kaon and a vector meson (physically a $K^*$ meson). Such a loop can couple to the pion. 
In turn this implies that the loop with the pion and the $\rho$ meson should be entirely 
insignificant at low pion masses, because it contributes with $M_\pi^4/(m_V^2 \, (4\pi F)^2)$ . 
We have checked that this is indeed the case (not shown here). 

To summarize, it is the not directly observable ratio of the physical to the bare pion mass
that receives a drastic correction from the loop with a ($K^*$) vector meson. That the result is numerically larger than the 
corresponding effect from the kaon and $\eta$-meson tadpole diagrams is interesting but not disturbing. 
Observable quantities might still agree with 
the rule of thumb that predicts that changes of the pion properties scale with powers of the pion mass. We reiterate our 
statement that we would have preferred to keep physical quantities constant when varying the bare pion mass. Yet, due to the 
complications discussed in the previous subsection this did not appear as a viable alternative within the present framework. 
With this qualitative understanding of the impact of the vector loops on the pion properties it should not be surprising 
that the effects for kaon and $\eta$ meson are also of comparable size. The effects are not very small, even for small 
bare pion masses.

\section{Summary} \label{sec:summary}

In this article, the influence of one-loop diagrams with dynamical vector mesons on masses and decay constants of pseudoscalar mesons is discussed. Thereby, the dependence on the bare pion mass as an input parameter is studied. Two studies are performed: First, the difference between static and dynamical vector mesons is examined. For all calculations, the difference turns out to be already significant for bare pion masses above approximately $250 \, \te{MeV}$. Second,  pure $\chpt$ calculations are compared to calculations involving vector-meson loops. Here, the calculations indicate that dynamical vector mesons are already important for low bare pion masses if the kaon mass is kept on its physical value. 

The studies performed in this article are based on a vector-meson Lagrangian which includes only a selected number of interactions terms. For studies with an extended vector-meson Lagrangian as suggested, \eg in \cite{Terschlusen:2012xw}, the influence of vector-meson loops on the renormalisation of the low-energy constants of $\chpt$ has to be determined first (cf.\ discussion in \cite{inf}). Equipped with such an information the properties of pseudoscalar mesons can be determined based on an even more realistic vector-meson Lagrangian. Yet already from the present work one can conclude that an analysis of the quark-mass dependence of lattice results might grossly underestimate the importance of vector mesons when such an analysis is based on pure $\chpt$.

%%
%%
%\begin{appendix}
%%
%%
%%
%%
%\end{appendix}
%%
%%

%\newpage
%\bibliographystyle{elsart-num}
%\bibliographystyle{plain}
%\bibliography{literature-t2.bib}
\bibliography{lit-PaperMass-2}

\end{document}